\documentclass[10pt, conference, letterpaper]{IEEEtran}
\IEEEoverridecommandlockouts

\usepackage{cite}
\usepackage[hidelinks]{hyperref}
\usepackage{tikz}
\usepackage[linesnumbered,ruled,vlined]{algorithm2e}     
\usepackage{amsmath, amssymb}
\usepackage{newtxtext,newtxmath}
\usepackage{filecontents}
\usepackage{subfigure}
\usepackage{diagbox}
\usepackage{booktabs}
\usepackage{graphicx}
\usepackage{circledsteps}

\usepackage{url}

\usepackage{color}
\usepackage{pifont}

\pgfkeys{/csteps/fill color=black}
\pgfkeys{/csteps/inner color=white}

\begin{document}

\title{Zeropod: Simplifying Datacenter Networking \\ with Future-Proof Zero-Buffer Packet Switches}

\author{
    Cong Liang$^{1}$, Jing Cheng$^{1}$, Mowei Wang$^{2}$, Yashe Liu$^{2}$, Zhenhua Liu$^{2}$, Yong Cui$^{1}$\\
    $^{1}$Tsinghua University\\
    $^{2}$Huawei Technologies Co., Ltd.
}

\maketitle

\begin{abstract}

    With the rapid growth of traffic volume in datacenter networks (DCNs), packet switches suffer from insufficient switching chip capacity and difficulties in transmission control, making it challenging to provide high goodput and low latency for emerging cloud applications. 
    
    We present Zeropod, a future-proof DCN architecture featuring simplified zero-buffer packet switches inside the point-of-delivery (pod). \textit{Within each pod}, traffic transmission is scheduled by a per-pod centralized scheduler for collision avoidance, enabling a highly simplified data plane, facilitating benefits like higher switching capacity and precise transmission control. \textit{Among the pods}, buffered Core switches work as barriers and relay inter-pod data, limiting the scope of centralized scheduling and thus simplifying the control plane. Zeropod combines host-level and flow-level scheduling for high performance with low overhead. Evaluation results show that Zeropod consistently performs better or equivalent to traditional buffered DCN, particularly regarding flow completion time (FCT). When accounting for the increased switching capacity due to the removal of buffers, its performance is further improved. Zeropod explores an extreme end of the design spectrum, and we hope it can encourage further exploration in the DCN community.
    
\end{abstract}

\section{Introduction}
\label{sec-introduction}

The continuous development of DCN applications~\cite{gao2016network, eashan23dbo, 10.1145/3695053.3731412, 10.1145/3651890.3672233} has increased the traffic volume in DCNs rapidly at roughly 100\% per year~\cite{ballani2020sirius} while also imposing high goodput and low latency expectations on DCNs. 
In contrast, the capacity of the CMOS-based switching chip, which is determined by the Serializer/Deserializer (SerDes) area, only doubles every 2 years~\cite{ballani2020sirius}, forming a gap between the traffic demand and the chip capability. Meanwhile, the packet buffers, which are used to absorb conflicts and support statistical multiplexing, become relatively short as the line rate increases, leading to packet drops or frequent PFC pauses and making the classical endhost-based transmission control ineffective~\cite{295507, bfc2022nsdi}.

These problems are further exacerbated in the post-Moore's Law era \cite{ballani2020sirius, moore2019another} and challenge the data transmission in traditional packet-switched DCNs, degrading their performance.
One line of work to address this issue introduces more functionalities to control the overwhelmed network more timely and precisely \cite{pyrrha, pred, li2019hpcc, bfc2022nsdi,bolt2023nsdi, 295507}. 
However, they do not reduce the burden of switching chips and may even complicate the chips with the advanced control logic, making it even harder for capacity growth.

Another line of work simplifies the switching chips by pushing network functionalities to the edges~\cite{zilberman2019stardust, jin2016your, firestone2018azure, li2018dumbnet, 10.1145/2486001.2491722}. For example, the forwarding table and its lookup logic, which may take 30\% of the switching chip area according to data from Intel \cite{intelhoti2020url}, can be offloaded to endhosts with source routing~\cite{jin2016your, li2018dumbnet}. 
The saved chip area can thus be used by other logic, such as to increase switching capacity with more SerDes area.
In the long run, this ``dumb network, smart edges'' trend has the potential to support the sustainable development of DCNs.

We explore the extreme end of this design spectrum and present Zeropod, a future-proof DCN architecture composed of pods with no in-network buffer that feature simplified, high-capacity, and zero-buffer packet switches.
Traffic traversing these switches is precisely scheduled by a per-pod centralized scheduler on an epoch basis to control both its transmission timing and path to avoid conflicts. To flexibly scale across the DCN, Zeropod interconnects multiple independent zero-buffer pods with buffered Core switches, which serve as coordination boundaries and isolate the centralized-controlled domains. 
High-capacity zero-buffer switching is deployed only within pods, aligning with DCN traffic patterns.

When incorporating zero-buffer switches into DCNs, scheduling (i.e., to decide the timing and path of data sending) is important to ensure conflict-free transmission and good flow-level performance. However, one DCN can have hundreds of thousands of machines \cite{roy2015inside} dynamically sending flows to each other, challenging the possibility of global flow-level scheduling.
To deal with this, by design, each Zeropod's centralized scheduler is in charge of the traffic inside its local pod instead of the whole DCN, lowering the scheduling burden. 
Moreover, Zeropod adopts a hierarchical scheduling approach: the scheduler grants data transmissions \textit{at host level} to avoid conflicts and ensure high goodput with low complexity, while each source host decides the priorities of local flow transmissions \textit{at flow level} once granted to optimize FCT for latency-sensitive mice flows. 
Such a ``hybrid'' method offloads part of the scheduling complexity from the centralized scheduler to endhosts, enabling flexible flow-level performance optimization with low overhead.

Although radical, Zeropod is still based on packet switching, meaning it can leverage the rich chip manufacturing and device deployment experience from traditional DCNs. 
Compared with circuit switching~\cite{shrivastav2019shoal,mellette2020expanding,ballani2020sirius,10.1145/3651890.3672273}, which also has no in-network buffer, Zeropod eliminates the need for path-level resource reservation, enabling flexible link-level hop-by-hop scheduling. Each link is scheduled independently and can be shared across multiple paths, potentially increasing network utilization. Also, packet header processing can be done by the packet switching chip, enabling capabilities like simple source routing.

We evaluate Zeropod with large-scale packet-level simulations. Results show that even without reusing the chip area saved by removing buffers, Zeropod achieves comparable or better performance than conventional DCNs. 
When the saved area is reallocated to SerDes logic to increase switching capacity, Zeropod significantly outperforms the conventional DCNs. With Zeropod, we hope to raise discussion in the community on whether the benefits of current approaches to designing DCNs are worth the complexity.

The main contributions of this paper are as follows:
\begin{itemize}
    \item We analyze the drawbacks of conventional buffered DCNs, which motivate zero-buffer packet switching (\textbf{\S\ref{sec-motivation}}).
    \item We present Zeropod, a future-proof DCN architecture for simpler control and higher capacity (\textbf{\S\ref{sec-overview}}).
    \item We propose a hybrid host-level (\textbf{\S\ref{sec-design-0}}) and flow-level (\textbf{\S\ref{sec-design-0-1}}) conflict-free scheduling mechanism.
 An in-band control plane is also designed (\textbf{\S\ref{sec-design-1}}).
    \item Through simulations, we show that Zeropod outperforms the conventional DCNs when reallocating the buffer area to SerDes logic for higher switching capacity (\textbf{\S\ref{sec-evaluation}}).
\end{itemize}

\section{Background \& Motivation}
\label{sec-motivation}

In this section, we first look back at the history of conventional packet switching and analyze why it is becoming insufficient for the sustainable development of recent DCNs. Then, we justify the effectiveness of zero-buffer packet switching, which motivates our design of Zeropod.

\subsection{Packet switching is becoming insufficient for DCNs}

\subsubsection{A brief history of packet switching}
\label{sec-packet-switching-history}
It has been around 60 years since packet switching was proposed \cite{baran1964distributed, davies1967digital}. When multiple flows compete for the same port, the switching chip absorbs the conflict with the on-chip buffer and statistically multiplexes the links. Although such dynamic link allocation requires the switching chip to be equipped with \textit{buffer} and \textit{computing power} to store and schedule the packets, it enables flexible and distributed coordination of switches, which is suitable for the \textit{large-scale internet} that is run by different administrators across the globe \cite{1455410}.

Therefore, as chip production technology matures to support the required buffer and computing power, packet switching gradually took over circuit-switched asynchronous transfer mode (ATM) and dominated the market. However, with the fast development of DCNs, which are held inside buildings by one administrator instead of across the globe, packet switching is now becoming insufficient for them.

\subsubsection{Packet switching struggles to fulfill the surging DCN traffic}
\label{sec-packet-switching-struggles}
Nowadays, an increasing number of applications are running in the DCNs. 
According to Microsoft and Google, the traffic volume within each DCN is expected to double annually, significantly outpacing the 2-year doubling rate of CMOS-based switching chips' capacity \cite{singh2015jupiter, ballani2020sirius}, potentially further slowing down in the post-Moore's Law era.
As a result, transmission control becomes more difficult.
The network is overwhelmed by the traffic, resulting in long queues in switch buffers, which leads to degraded performance, increased costs, and higher power consumption, ultimately making packet switching unsustainable for the future.

Luckily, unlike the internet, a single DCN is often housed in several buildings and has a single administrator. Innovative designs can be applied, with recent simplified DCNs being an example \cite{zilberman2019stardust, jin2016your, firestone2018azure, li2018dumbnet, 10.1145/2486001.2491722}. 
This motivates us to rethink the network architecture suitable for DCNs.

\subsection{Zero-buffer packet switching as a sustainable option}
\label{zero-buffer-packet-switching-as-a-sustainable-option}

Although ``dumb networks, smart edges'' has been a popular idea in the networking community from the very beginning~\cite{10.1145/52324.52336, 10.1145/357401.357402}, there are few works that have explored the possibility of removing packet buffers from packet switches.
We argue that by doing so, the mismatch between the traffic volume and the switching capacity can be alleviated, contributing to a more sustainable DCN.

\subsubsection{Higher switching capacity and lower cost}
\label{sec-zero-buffer-packet-switching-higher-switching-capacity-and-lower-cost}

The switching capacity of a chip (i.e., the maximum throughput it can support) is determined by the chip area allocated to SerDes logic. However, in the post-Moore’s Law era, achieving further power and area scaling is increasingly difficult~\cite{ballani2020sirius, moore2019another}, making it harder to expand SerDes and thus limiting capacity.

Simplifying the switch with zero buffering and zero forwarding (i.e., using source routing instead of forwarding tables) can free up substantial chip area for additional SerDes, providing a sustainable way for continuous scaling.
According to Intel's data, in modern switch chips, the packet buffer and the control logic of forwarding and buffering occupy about the same amount or more of die size with SerDes \cite{intelhoti2020url}, which can be translated to a 2$\times$ SerDes area if the buffer is removed.
When SerDes do not use the saved area, the simplified chips will save die size and power consumption, helping to cut the overall cost of DCN. This has the potential to reduce the reliance on CMOS technology of chip manufacturing, implying that switching chips made using the $N^{th}$ generation technology (e.g., 5nm) could achieve comparable performance to those made with the $N+1^{th}$ (e.g., 3nm).

While using a smaller buffer can also reduce chip area, the benefit is limited by the fixed cost of buffer control logic.
Our switch chip provider estimates that the area of the buffer control logic is approximately fixed. When using the current normally-sized buffer as a reference, this fixed control logic area can consume twice as much die size as the buffer itself, which is quite considerable. As long as the buffer exists, no matter its size, the fixed area of control logic will remain the same and cannot be released. Moreover, as line rates continue to increase (e.g., from 10Gbps~\cite{roy2015inside} to 400Gbps link~\cite{10.1145/3651890.3672233}), the threshold for a ``small'' buffer also goes up. Eventually, we will face a similar control complexity with ``zero'' buffer switches.

\subsubsection{Elimination of buffer management}
\label{sec-zero-buffer-packet-switching-elimination-of-buffer-management}

In buffered DCNs, switch buffer management is a challenging task, especially when the increase of the buffer size cannot keep up with that of the line rate \cite{295507, bfc2022nsdi}. On one hand, researchers have to use more complex transmission control mechanisms for a lower queue length in switches~\cite{pyrrha, pred, li2019hpcc, bfc2022nsdi,bolt2023nsdi, 295507}, which do not reduce the burden of switching chips and may even make it harder for their scaling if switch-side modification is needed, reserving less chip area for the buffer in turn. On the other hand, it is difficult to adjust the buffer-related parameters. For example, the tunable parameter $\alpha$ for dynamic threshold in shared buffer switches~\cite{664262} and the classic ECN threshold used in congestion control schemes like DCTCP\cite{alizadeh2010data} as well as DCQCN~\cite{zhu2015congestion} are all buffer-related parameters that directly affect network performance but are hard to tune~\cite{abm2022addanki,acc2021yan,meta2022imc}.
When switch buffers are removed, the above buffer management problem is naturally eliminated.

In terms of small buffers, the difficulty in buffer management is not eased. Instead, since the available buffer resource is limited, the damage of inappropriate management will become even more severe~\cite{abm2022addanki}.

\subsubsection{More flexible scheduling than circuit switching}
\label{sec-zero-buffer-packet-switching-more-flexible-scheduling-than-circuit-switching}

Circuit-switched DCNs~\cite{shrivastav2019shoal, mellette2020expanding, ballani2020sirius, 10.1145/3651890.3672273, 10.1145/3651890.3672222} also eliminate in-network buffers. However, they require pre-allocated end-to-end circuit paths that must remain reserved for the entire duration of the transmission and can only be released upon completion. 
In contrast, for packet switching, the links can be scheduled independently and can even be shared across multiple paths, potentially achieving better link utilization.

\section{Zeropod's building blocks}
\label{sec-overview}

\begin{figure}[t]
 \centering
 \includegraphics[width=0.48\textwidth]{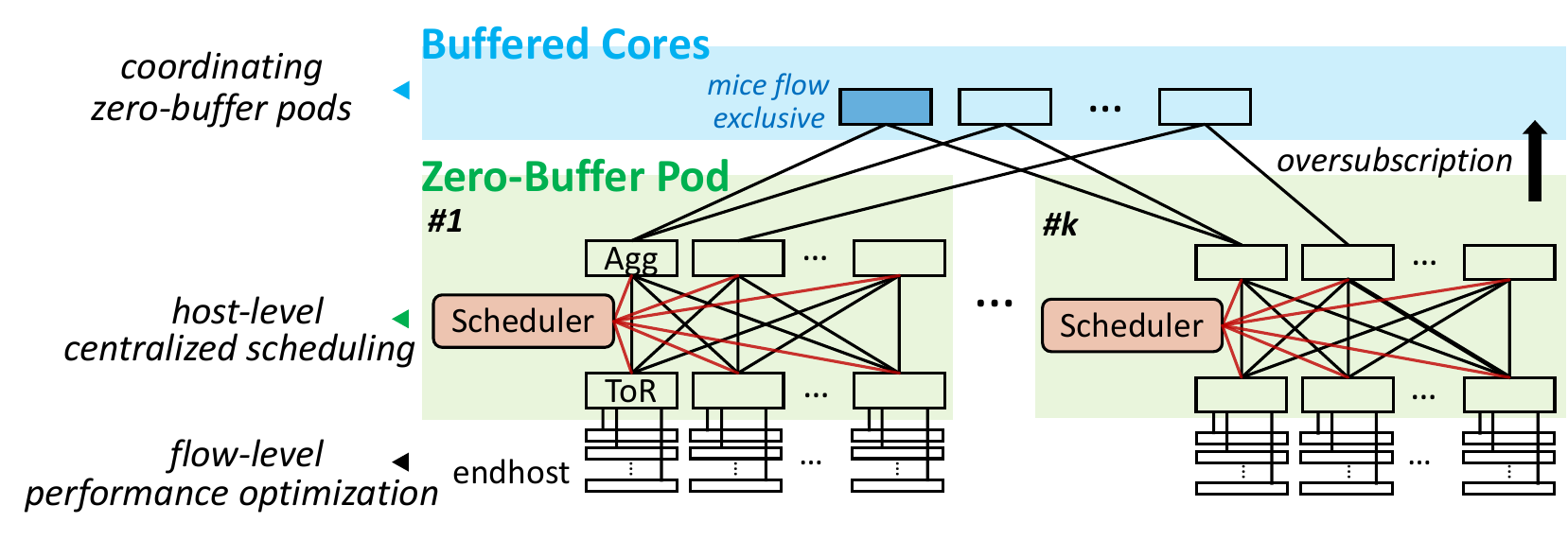}
 \caption{Zeropod architecture using a Fattree-like topology with $k$-port intra-pod switches. Each pod contains $k^2/4$ endhosts, with $k^3/4$ endhosts in total.}
 \vspace{-0.1in}
 \label{fig-architecture}
\end{figure}

\subsection{Zeropod overview}

As shown in Fig.~\ref{fig-architecture}, Zeropod adopts a Fattree-based topology \cite{fattree}, consisting of several fully-provisioned zero-buffer interconnected by oversubscribed, buffered Core switches to interconnect the pods. All links work at the same rate. 
Intra-pod transmissions are \textbf{synchronously} and precisely controlled by the local centralized scheduler to avoid conflict. 
For inter-pod transmissions, the source-side scheduler and the destination-side scheduler collaborate with Cores respectively, \textbf{asynchronously} supporting DCN-scale traffic.

Inside one pod, buffer-less Top-of-Rack (ToR) and Aggregation (Agg) switches form a zero-buffer domain. One scheduler is directly connected to all ToRs and Aggs in this pod to enable efficient in-band delivery of control messages.
Each endhost is equipped with a smartNIC that precisely controls its sending behavior. 
For the ease of scheduling, Zeropod cuts or aggregates packets for the same source-destination pair into fixed-size cells. Inside each pod, time is divided into fixed-length epochs and is synchronized among the endhosts, with one epoch being the scheduling granularity. Each epoch comprises multiple cells from the same set of source-destination pairs. 
Every time new traffic arrives, the endhost will send a request-to-send (RTS) to the scheduler in the same pod, specifying the destination and the newly arrived volume of data for this destination. The data will be held until the schedule (SCHD) is received. 
Data will be sent at the exact time via the given path indicated by SCHD.

Note that both time synchronization and centralized scheduling are limited to single pods. For inter-pod transmissions, buffered Core switches operate as barriers to isolate the pods. 
As such, the Cores distributedly coordinate the zero-buffer pods, expanding Zeropod to DCN scale.

\subsection{Intra-pod: Zero-buffer domain with centralized control}

\subsubsection{Switching techniques}
\label{sec-intra-pod-switching-techniques}
On endhosts' side, Zeropod uses per-destination virtual output queues (VOQs) to hold packets arriving from the upper-layer stacks. At the NIC, for each VOQ, packets are cut or aggregated into fixed-length cells before sending, improving the efficiency of scheduling and switching. 
Each cell contains a small header, which includes a sequence number that allows the cells to be reassembled back into packets at the destination.

Zeropod uses source routing techniques to remove forwarding tables and complex forwarding logic from switches, achieving zero-forwarding for further simplification~\cite{li2018dumbnet, jin2016your}. 
The scheduler makes forwarding decisions for data cells. Each NIC holds a source routing table that interprets scheduling results to source routing information in cell headers, which indicate the hop-by-hop forwarding port of switches along the path. 
At each hop, the switch parses the routing information and forwards the cell to the corresponding port.

\subsubsection{Host-level pipelined centralized scheduling (\S\ref{sec-design-0})}
\label{sec-intra-pod-host-level-pipelined-centralized-scheduling}
To reduce the computational overhead, Zeropod schedules at the coarse-grained host level instead of the flow level.
The scheduler processes each RTS, updates the traffic matrix, and ensures collision-free transmissions using an iterative scheduling algorithm. 
An SCHD message is then sent back to the sender, indicating the allocated transmission time, destination, data volume, and a non-conflicting transmission path.

To fully utilize the network and eliminate waste of epochs, the exchange of RTS and SCHD is pipelined, where one endhost sends an RTS in one epoch and receives the corresponding SCHD in a later epoch.
After sending an RTS cell to the scheduler, instead of getting stuck and waiting for the scheduling result to return, the sender continuously sends RTS cells as well as data cells in the following epochs, making full use of the fabric.
After receiving SCHD, the sender then executes the scheduling results at the corresponding time.

\subsubsection{Flow-level hybrid performance optimization (\S\ref{sec-design-0-1})}
\label{sec-intra-pod-flow-level-hybrid-performance-optimization}
Based on the host-level scheduling, to improve flow-level latency performance, we integrate a hierarchical hybrid flow optimization mechanism into Zeropod. 
Specifically, we attach priorities to RTS to favor endhosts with mice flows. Additionally, once scheduled, endhosts prioritize mice flows when generating cells. 
To further reduce FCT, we also introduce an optimistic sending mechanism that allows mice flows to bypass scheduling delays if possible.

\subsubsection{In-band control plane (\S\ref{sec-design-1})}
\label{sec-intra-pod-in-band-control-plane}
Zeropod needs to frequently exchange control messages between the scheduler and the senders in the corresponding pod. It achieves this through the same network fabric for data transmissions. 
The basic idea is to selectively reserve gaps between data cells in each epoch, so that RTS and SCHD can be inserted to avoid collision among themselves or with data cells. 
To determine the sending time and the position to insert, we use a predefined scheduler-less scheme to avoid incurring extra coordination efforts to senders and the scheduler.

\subsection{Inter-pod: Asynchronous two-stage transmission}

To enable inter-pod communications while isolating the zero-buffer pods to avoid the complexity of global scheduling and time synchronization, Zeropod connects the pods with buffered Core switches that are used to relay traffic. 
There are two stages of scheduling that happen asynchronously for inter-pod traffic: one stage in the source pod, and the other in the destination pod, with the Core switches acting as intermediate nodes and coordinating the two stages.
They behave like endhosts with almost the same functionalities, and are shared by all pods in a load-balanced manner for traffic relay. Cells are first sent up to the Cores by the source-side scheduler with one request-grant process, and then sent down to the destination endhost by the destination-side scheduler with another request-grant process. 

In Core switches, each port maintains per-endhost VOQs for the pod it connects to, and needs to track multiple clocks, one for each connected pod. 
To prevent queue buildup and buffer overflow at Core switches, Zeropod incorporates a backpressure mechanism (\S\ref{sec-design-0}). Furthermore, to mitigate HoL blocking and improve inter-pod FCT, it assigns exclusive Cores for mice flows (\S\ref{sec-design-0-1}).

\section{Design Details}
\label{sec-design}

\subsection{Host-level pipelined centralized scheduling}
\label{sec-design-0}

Zeropod utilizes a host-level traffic scheduling approach to control intra-pod and inter-pod traffic simultaneously. 
To ease the difficulty of hardware implementation and make the centralized scheduler possible, we design an iterative scheduling algorithm that comprises several phases and could be distributed to multiple processing cores, enabling the algorithm's pipelined execution. In this section, we first show the goal and workflow of the scheduling task, and then present the algorithm run by the scheduler.

\subsubsection{The goal and workflow of scheduling}
\label{sec-design-goal_analysis}
Even though Zeropod uses a fully-provisioned two-layer Clos topology inside each pod, due to the existence of one-directional inter-pod traffic that goes up or down, the scheduling problem is not a non-blocking one. 
There are three kinds of traffic that the scheduler needs to manage.

\begin{itemize}
    \item Intra-pod \textbf{E2E} (Endhost-to-Endhost) traffic with its source and destination inside one pod. The E2E traffic is routed up to Agg and then routed down for effortless scheduling of zero-queueing transmission in the bufferless fabric.
    \item Up-facing \textbf{E2P} (Endhost-to-Pod) traffic sent from endhosts to Cores for intermediation. It is the first stage of inter-pod traffic, and the scheduler in the source pod needs to allocate one available ToR-Agg-Core path to it.
    \item Down-facing \textbf{C2E} (Core-to-Endhost) traffic sent from Cores to endhosts in the destination pod. It is the second and final stage of inter-pod traffic, and the scheduler in the destination pod needs to verify the availability of the required Core-Agg-ToR path.
\end{itemize}

Each scheduler maintains three traffic matrices accordingly. 
Aggregated sizes of data to send are stored in these matrices. The RTS cells inform the scheduler of newly arrived data so that the scheduler can update its traffic matrices. When an SCHD is given, the matrices are also updated to subtract the granted data size.

\subsubsection{Buffer control of Core switches}
\label{sec-design-0-2}
To relay inter-pod traffic, each port of a Core switch is connected to a pod, where E2P data is temporarily buffered and then sent as C2E traffic to the destination pod, utilizing a similar RTS-SCHD-based mechanism with endhosts. 
Since the Core switches for data relay are selected independently by schedulers from different pods, queue buildup and buffer overflow may happen. 
To solve this, we introduce a PFC-like backpressure mechanism to Cores.
When the queue length of an output port exceeds a predefined threshold, backpressure signals are tagged in RTS and broadcast to schedulers in all pods. This informs the schedulers that traffic relayed to the corresponding pod through this specific Core is not allowed, prompting them to adjust decisions accordingly. 
We test buffer occupation of Cores and the impact of threshold values in evaluation (\S~\ref{sec-evaluation}).

\subsubsection{The scheduling algorithm}
\label{sec-design-0-3}

To achieve a high scheduling throughput, we design an iterative scheduling algorithm (Algorithm~\ref{alg:iterative-scheduler}) which is \textbf{composed of iterations of four phases and can be implemented with multiple processor cores in a pipelined manner}. 
The number of processor cores needed by each phase can be chosen according to the network scale and the scheduling delay expectation, so that each phase is fast enough and aligns with other phases to form a pipeline.

\noindent \Circled{\textbf{1}} \textbf{Request: Path selection}. For one demand, the scheduler randomly requests the specific idle links needed to form a path, based on the switch and endhost port states provided as input.
If it is E2P traffic, the requested Core must have no backpressure signal.

After this phase, one port might be requested by multiple demands, resulting in conflicts. The request list is passed to later phases to eliminate such conflicts.

\noindent \Circled{\textbf{2}} \textbf{Grant 1: Resolving switch port conflicts}. 
As we analyzed in \S\ref{sec-design-goal_analysis}, for a given flow, once the Agg is chosen, the required ports at ToRs and Cores are naturally determined. We thus focus on Aggs when resolving switch port conflicts.
For each Agg switch with multiple requests, we allocate its available ports to requests. The results will be passed to later phases along with the updated port occupation state.

\noindent\Circled{\textbf{3}} \textbf{Grant 2: Resolving sender-side conflicts}. 
Approve one grant from each sender's grant list.
For the unapproved grants, the corresponding switch ports are restored as unoccupied. 
The approved grants, along with the port occupation state, are passed to the next phase.

\noindent \Circled{\textbf{4}} \textbf{Accept: Resolving receiver-side conflicts}. 
Finally, choose one grant to accept from each receiver's grant list. 
For the unapproved grants, the corresponding ports are also restored as unoccupied.
Now the scheduling result of this iteration is obtained, including the demand and its assigned path. 

Although more iterations could be done to improve scheduling efficiency, this would also increase the scheduling delay and thus FCTs.
Trade-offs should be made for different workloads: bursty workloads that are latency-sensitive prefer fewer iterations, while bulk workloads prefer the opposite.
We set the iteration number to 6 in our evaluation (\S~\ref{sec-evaluation}).

\SetKwInput{KwIn}{Input}
\SetKwInput{KwOut}{Output}
\SetKw{KwRet}{return}
\newcommand{\Call}[2]{\textsc{#1}(#2)}

\begin{algorithm}[t]
\caption{Zeropod's Iterative Scheduling}
\label{alg:iterative-scheduler}
\KwIn{Demands $\mathcal{D}$; port states $\mathcal{S}$; max iterations $I$}
\KwOut{Set of scheduled demands and paths $\mathcal{O}_{\rm sched}$}
$\mathcal{O}_{\rm sched} \leftarrow \varnothing$\;

\For{$iter \leftarrow 1$ \KwTo $I$}{ 

$Req \leftarrow \varnothing$; $G_1 \leftarrow \varnothing$; $G_2 \leftarrow \varnothing$; $Accept \leftarrow \varnothing$\;

  \tcp{{\rm\footnotesize\Circled{\textbf{1}}} Request (path selection)}
  \ForEach{$d \in \mathcal{D} \setminus \mathcal{O}_{\rm sched}$}{

    \ForEach{$path \in \Call{SelectIdlePaths}{d, \mathcal{S}}$}{
        \If{($d$ is E2P \textbf{and} \Call{HasBackpressure}{$path$})}{
            continue\;
        }
        $Req \leftarrow Req \cup \{(d, path)\}$\;
    }
  }
  \tcp{{\rm\footnotesize\Circled{\textbf{2}}} Grant 1 (switch-port conflicts)}
  \ForEach{Agg switch $agg$ with requests in $Req$}{
    $\{(d, path), ...\} \leftarrow \Call{AllocatePorts}{Req[agg]}$\;
    $G_1 \leftarrow G_1 \cup \{(d, path), ...\}$\;
    $\mathcal{S} \leftarrow$ \Call{MarkOccupied}{$\{(d, path), ...\}, \mathcal{S}$}\;
  }
  \tcp{{\rm\footnotesize\Circled{\textbf{3}}} Grant 2 (sender-side conflicts)}
  \ForEach{sender $s$ with grants in $G_1$}{
    $(d, path) \leftarrow \Call{AllocatePort}{G_1[s]}$\;
    $G_2 \leftarrow G_2 \cup \{(d, path)\}$\;
    $\mathcal{S} \leftarrow$ \Call{ReleasePorts}{$G_1[s] \setminus \{(d, path)\}, \mathcal{S}$}\;
  }
  \tcp{{\rm\footnotesize\Circled{\textbf{4}}} Accept (receiver-side conflicts)}
  \ForEach{receiver $r$ with grants in $G_2$}{
    $(d, path) \leftarrow \Call{AllocatePort}{G_2[r]}$\;
    $Accept \leftarrow Accept \cup \{(d, path)\}$\;
    $\mathcal{S} \leftarrow$ \Call{ReleasePorts}{$G_2[r] \setminus \{(d, path)\}, \mathcal{S}$}\;
  }
  $\mathcal{O}_{\rm sched} \leftarrow \mathcal{O}_{\rm sched} \cup Accept$\;
}
\KwRet{$\mathcal{O}_{\rm sched}$}
\end{algorithm}

\subsection{Flow-level performance optimization}
\label{sec-design-0-1}

With host-level scheduling, Zeropod achieves high-goodput conflict-free transmission in zero-buffer domains.
However, the FCT of flows, especially mice flows, is also crucial for the application's performance.  
To prioritize mice flows, when endhosts get scheduled and generate data cells to send, they consider their local flow information with PIAS-like information-agnostic multi-level feedback queue \cite{bai2015information} in per-destination VOQs and send mice flows heading for the same destination first. 
We further explore several designs to optimize the FCTs for mice flows.

\subsubsection{Feeding top-$k$ priority information to the scheduler}
\label{sec-design-0-1-1}

Other than prioritizing mice flows locally at each sender, Zeropod can also prioritize senders with mice flows at the scheduler side.
A straightforward solution schedules traffic with fewer aggregated data first. However, this is not enough since the aggregated data size in the scheduler's traffic matrix can hide the actual flow-level demands. 

To fix this, we introduce priority information in RTS cells to expose mice flow information to the scheduler. For each endhost, when generating RTS, except for the newly arrived data size, it also marks its priority, which contains the endhost's current top-$k$ destinations with the shortest flows along with the sizes of these mice flows, which can be estimated from the endhost's local PIAS \cite{bai2015information} queue. 
When deciding the scheduling order for source-destination pairs, if mice flow size is given, the scheduler will consider the mice flow size instead of the aggregated data size. 
Here $k$ is a variable and can be tuned. With a larger $k$, the scheduler can approximate a better mice-flow-first scheduling, but the RTS control overhead will also increase. 
Our evaluation (\S~\ref{sec-intra-pod-evaluation}) shows that a relatively small $k$ is enough to bring considerable FCT optimization.

\subsubsection{Alleviating HoL blocking at Core switches}
\label{sec-design-0-1-2}
The previous optimizations work fine for intra-pod traffic. However, for inter-pod traffic, considering the flow size in DCNs exhibits a heavy-tailed distribution where elephant flows occupy the majority of the bytes \cite{montazeri2018homa}, inside Core switches, mice flow traffic may be HoL blocked by bulk elephant flow traffic after getting into the same FIFO queue. This will damage their FCT even if endhosts always send them first.

Zeropod alleviates this problem by adding extra Core switches and fully utilizing the flexibility of centralized scheduling to ensure that the extra Cores are exclusively used for traffic containing mice flows.
These extra Cores reduce the queueing delay that mice flows may experience. Evaluation (\S~\ref{sec-inter-pod-evaluation}) shows one such extra Core can reduce the FCT of mice flows by more than 30\% under high load.

\subsubsection{Bypassing scheduling delay with optimistic sending}
\label{sec-design-0-1-3}
Zeropod's centralized control introduces scheduling delays. For mice flows, this delay will severely affect their completion time, especially when the network is lightly loaded and mice flows could have reached their destinations without collision even if they randomly choose the paths.

To solve this, we use an optimistic sending mechanism.
After sending RTS, before receiving the corresponding SCHD, 
endhosts try to optimistically send unscheduled data cells from mice flows via a randomly chosen path if no scheduled data is being sent.
The unscheduled cells are less prioritized and will be dropped on collision with scheduled cells in the fabric, so that \textbf{the reliability promise for scheduled cells is never compromised}. If all cells of a mice flow successfully reach the destination, then its FCT will be vastly reduced. 
Later, the sender will send the data as scheduled cells again after receiving the SCHD. 
Note that the transmission of optimistic cells is transparent to RTS-SCHD-based scheduling. As a result, it \textbf{only consumes the bandwidth that is otherwise unused and will NOT introduce bandwidth waste}.

\begin{table}[t]
    \centering
    \caption{Key notations in the analysis.}
    \label{tab-symbols}
    \begin{tabular}{|c|c|}
        \hline
         & \textbf{Description} \\ 
        \hline
        $k$ & Number of ports per switch in the Fattree topology \\
        \hline
        $m$ & Number of endhosts under each ToR, $m=k/2$ \\
        \hline
        $t$ & Time needed to send a cell at line rate \\
        \hline
        $\lambda$ & \#Cells generated per endhost during $t$ ($0 < \lambda < 1$) \\
        \hline
        $G$ & Expected network load of one endhost  \\
        \hline
        $S$ & Expected goodput of one endhost \\
        \hline
        $p_i$ & Prob. of one endhost generating $i$ cells during a cell time $t$ \\
        \hline
        $P$ & Prob. of successful delivery of one unscheduled cell \\
        \hline
        $H$ & Prob. of one unscheduled cell choosing a path w/o collision \\
        \hline
        $^{iR, oR}$ & Superscripts to indicate ``intra-rack'' and ``inter-rack'' \\
        \hline
    \end{tabular}
\end{table}

Here, we theoretically model the performance of optimistic sending. To get \textit{the goodput of unscheduled cells that can reach the destination without collisions}, we build a simple model where Zeropod's zero-buffer fabric is treated as a shared medium with a layered topology. 
We summarize the key notations used in the following analysis in Table \ref{tab-symbols}. To simplify the analysis, \textit{we assume that there is only unscheduled intra-pod traffic in the network}. 
Consider a pod in a $k$-port Fattree with $m=k/2$ endhosts under each ToR. Each endhost generates unscheduled cells following a Poisson distribution. 
In such a pod, each endhost generates $i$ cells in a cell time $t$ at probability $p_i= \frac{G^i e^{-G}}{i!}$ with $G=\lambda t$ being the expected network load for this endhost, resulting in a goodput of
\begin{equation}
S=GP
\label{eq-goodput}
\end{equation} 
where $P$ is the probability of successful delivery of one given unscheduled cell.

To get $P$,
we consider two scenarios. For the first scenario, there is no other endhost sending cells within a total time of $2t$ before and after the time when the given cell is sent, and thus collision will not happen. For the second scenario, during this $2t$ time, other endhosts send $i$ cells. However, their transmission paths are all different from that of the given cell, and we represent its probability as $H^i$, with $H$ being the probability that another cell does not conflict with the path of the given cell. Then we have $P = \sum_{i=0}^{\infty} p_iH^i \cdot \sum_{i=0}^{\infty} p_iH^i = e^{2G(H-1)}$. By considering intra-rack and inter-rack cells separately, we further have
\begin{equation}
P=e^{2G^{iR}(H^{iR}-1)}  e^{2G^{oR}(H^{oR}-1)}
\label{eq-P}
\end{equation}
Regarding $G^{iR}$ and $G^{oR}$, in the Fattree topology, the intensity of the aggregated Poisson flow from the same and different source racks with the given cell is
\begin{equation}
G^{iR}=mG \quad \text{and} \quad G^{oR}=m(m-1)G
\label{eq-G}
\end{equation}
respectively. Then we consider $H^{iR}$ and $H^{oR}$, which are the probabilities of one unscheduled cell, from an endhost in the same rack or outside the rack, respectively, choosing a ToR$\rightarrow$Agg$\rightarrow$ToR path that does not collide with the given cell. On the layered Fattree topology, we have
\begin{equation}
H^{iR} = \frac{m-1}{m}(\frac{m-1}{m}+ \frac{1}{m}\frac{m-1}{m})
\label{eq-H-iR}
\end{equation}
\begin{equation}
H^{oR} =\frac{1}{m}\frac{m-1}{m}+\frac{m-1}{m}(\frac{m-1}{m}+\frac{1}{m}\frac{m-1}{m})
\label{eq-H-oR}
\end{equation}

Substituting Equation \ref{eq-G}, \ref{eq-H-iR}, and \ref{eq-H-oR} into Equation \ref{eq-P}, we have \textbf{the probability of no collision}
\begin{equation}
P = e^{2G^{iR}(H^{iR}-1)}  e^{2G^{oR}(H^{oR}-1)}
=e^{2G(\frac{2}{m}-3)}
\label{eq-P-final}
\end{equation}
where $\lim_{m \to \infty} P = e^{-6G}$ is its asymptotic lower bound. Equation \ref{eq-goodput} and \ref{eq-P-final} yield \textbf{the goodput of a single endhost}
\begin{equation}
S = Ge^{2G(\frac{2}{m}-3)}
\end{equation}
which achieves the maximum value $S=\frac{1}{(6-\frac{4}{m})e}$ when $G=\frac{1}{6-\frac{4}{m}}$. 
This means for a $k=32$ Fattree, when the network load is low (say 10\%), more than 55\% cells can reach their destinations without collisions, yielding a considerable improvement on FCT. In \S~\ref{sec-intra-pod-evaluation}, we verify the effectiveness of this mechanism through simulations.

\subsection{The in-band control plane}
\label{sec-design-1}

Zeropod requires the endhosts and Cores to exchange RTS and SCHD with the scheduler in every epoch. 
To ease the operational cost, we reuse the same zero-buffer fabric for exchanges of control messages. 
This requires Zeropod to avoid potential collisions caused by control cells. 
Taking scheduler-endhost as an example, at the sender side, RTS cells from different endhosts under the same ToR may collide with each other at the ToR's egress port towards the scheduler if not staggered. 
Similarly, at the receiver side, the SCHD may collide with data cells at the ToR's egress port to the endhost. The scheduler-Core case faces similar problems.

We design a predefined gap-filling rule to stagger the transfer of control cells explicitly.
All RTS and SCHD in Zeropod share the same size.
Inside each epoch, the sender sends one RTS at the $i$th control slot and keeps the $j$th control slot empty for future SCHD, where $i$ and $j$ are the sender's and receiver's relative position of all the endhosts under the same ToR/Agg, respectively, staggering the cells. 
Since the scheduling results (i.e., when to send and to whom) are known by both the scheduler and the sender, the position of these control slots can be determined locally without extra coordination.

To ensure that every endhost can communicate with the scheduler once per epoch, the number of possible control slots, $N+1$ where $N$ is the number of data slots in an epoch, is set to be at least the number of endhosts under a ToR/Agg (i.e., $N+1 \geq k/2$). 
The total size of the two control cells should be smaller than the size of a single data cell; otherwise, the gaps at adjacent positions will overlap.

\subsection{An example workflow}

\begin{figure}[t]
    \centering
    \includegraphics[width=0.46\textwidth]{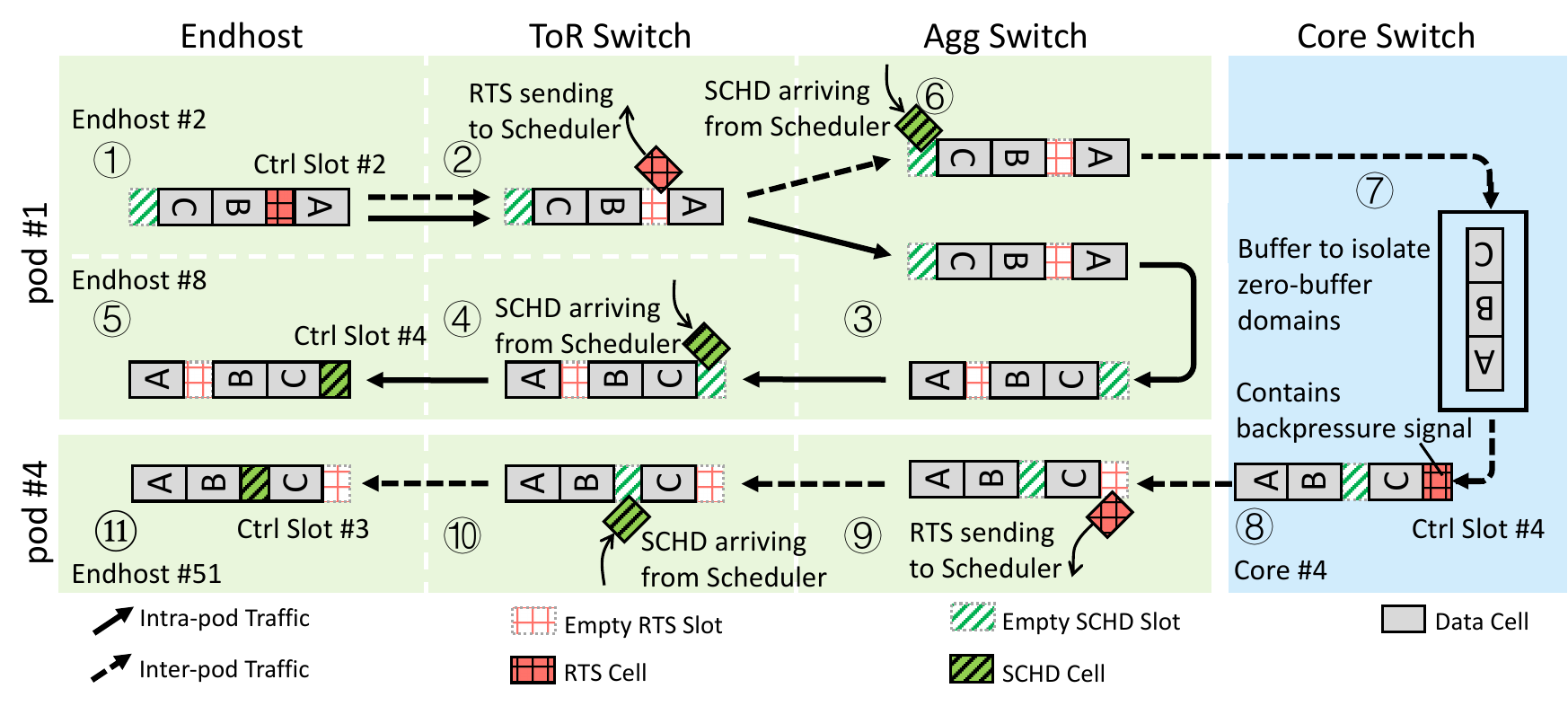}
    \vspace{-0.1in}
    \caption{Workflow when using a $k=8$ Fattree, and an $N=3$ data cell epoch.
    (1)-(5) depicts for the intra-pod while (1)(2)(6)-(11) for the inter-pod traffic.}
    \vspace{-0.2in}
    \label{fig-workflow}
\end{figure}

Fig.~\ref{fig-workflow} shows an example workflow of one epoch's traffic. For simplicity, we do not show optimistic sending. Note that since C2E traffic reaches Agg switches earlier due to fewer hops, epochs in Cores are precisely delayed to align C2E with E2E that share the same Agg downlink ports to avoid conflicts.

\subsection{Deployment considerations}
\label{sec-practical-concerns}

\noindent\textbf{Time synchronization.}
Endhosts belonging to a pod are required to be time-synchronized to enable Zeropod's epoch-based slotted operation. Different pods can operate asynchronously since Core switches have buffers to absorb the time difference between pods. 
In principle, Zeropod can be built with any synchronization protocols~\cite{lee2016globally, kannan2019precise}.
The scheduler works as the main clock, and the endhosts and Core ports in the same pod synchronize their clock with it using cyclic RTS and SCHD. Dummy control messages thus are exchanged even when the system is idle. Because the underlying network has no buffer, the one-way delay in Zeropod is thus deterministic and in turn facilitates precise time synchronization. 
To absorb synchronization error, we can set a ``guardband'' between epochs and around control slots, which is a common practice in time-synchronized systems \cite{ballani2020sirius, shrivastav2019shoal, mellette2020expanding}.

\noindent\textbf{Failure detection and recovery.}
Zeropod uses cyclic RTS and SCHD (or dummy ones when the system is idle) to detect link-down failures. 
The links belonging to the paths between the scheduler and senders can be detected as a failure if RTS/SCHD does not arrive on time. For other links, the sender can report a flag in the RTS to the scheduler, indicating whether it received any scheduled cells or not. 
The scheduler then uses this information to compare with previously stored scheduling results to infer whether and where a failure occurs. TCP-like ACK-based reliability mechanism can also be introduced to the upper layers in case of link flapping errors. To deal with scheduler failure, Zeropod can prepare one backup scheduler and use coordination services like ZooKeeper \cite{267345} to synchronize states among schedulers for a smooth scheduler transition.

\section{Evaluation}

\label{sec-evaluation}

In this section, we evaluate the performance of Zeropod based on a packet-level network simulator YAPS~\cite{PHOST}. 
We first evaluate Zeropod in a single pod, and then shift our focus to the DCN scale. 
Our results show that Zeropod consistently performs better or equivalent to conventional buffered DCN. 
Its performance gets better when we translate the saved chip area into increased switching capacity.
This demonstrates the effectiveness of our design.

\subsection{Simulation setup}

\noindent \textbf{Network topology.} We set the topology as Fig. \ref{fig-architecture}. 
For single-pod experiments, we take one pod from a Fattree with $k$ = 32, which contains 16 ToRs, 16 Aggs, and 256 endhosts. 
For multi-pod experiments, we use a Fattree with $k$ = 8, comprising 4 ToRs, 4 Aggs, and 16 endhosts inside one pod, and 128 endhosts in total. 
Unless specified, we use 7 Core switches to connect all pods in multi-pod experiments, with an oversubscription ratio of 16:7 compared to a standard Fattree. 
Per-hop propagation delay is all set to 1 $\mu$s. 
The linerate is set to 100 Gbps by default. Since Zeropod essentially saves the switching chip area (\S\ref{zero-buffer-packet-switching-as-a-sustainable-option}),
we reallocate the saved area to SerDes and translate it to a 2$\times$ switching capacity, enlarging the linerate to 200Gbps. We refer to it as Zeropod (2$\times$).

\noindent\textbf{Traffic workloads.} Unless specified, we adopt a widely used datacenter traffic workload WebSearch~\cite{alizadeh2010data}, using a Poisson arrival process for specified target network loads each containing 100,000 flows. All the flows follow an all-to-all pattern. The source and the destination of a flow are chosen uniformly in single-pod experiments. 
To simulate the locality of actual traffic \cite{benson2010network, kandula2009nature, roy2015inside}, in multi-pod experiments, the intra-pod traffic takes 80\% of the load while the inter-pod traffic takes 20\%. 
We also test Zeropod with typical microbenchmarks, such as incast and collective communication.

\noindent\textbf{Baselines.} We compare Zeropod with an ideal solution as well as traditional buffered DCNs, which use DCTCP~\cite{alizadeh2010data}. The ideal solution, described in~\cite{alizadeh2013pfabric}, leverages global flow information, schedules flows in non-decreasing order of remaining flow sizes, and allows shorter flows to preempt the longer ones. 
Although not practical, we use it to serve as an upper bound of performance. Both baselines use the same topology as Zeropod but use buffered switches.

\noindent\textbf{Parameter setup.}
For Zeropod, we set the round of iterative scheduling to 6. The scheduling delay is 15 $\mu$s for single-pod experiments, which have 256 endhosts per pod, and is 5 $\mu$s for multi-pod experiments with 16 endhosts per pod.
Each data cell contains a 512B payload and a 40B header.
The size of one control cell, including RTS and SCHD, is 40B. The optimistic sending mechanism is only enabled for flows shorter than 5 data cells. The buffer size of Core switches is set to 32MB.
We assume perfect time synchronization. For the traditional DCNs, each switch port contains a 512 KB buffer, resulting in a total switch buffer of 16 MB, which is derived from real device configurations. We set the ECN threshold as 1/7 bandwidth-delay product according to the DCTCP paper \cite{alizadeh2010data}.
Zeropod uses an epoch length of 16 cells in the single-pod experiments and 32 cells in the multi-pod experiments, and we omit their parameter sensitivity experiments due to space limitations. It piggybacks top-16 destinations with mice flow info in the RTS, and sets the backpressure threshold of each per-port buffer for Core switches to 320 data cells. We also explore other setups in the following experiments.

\noindent\textbf{Metric used.} We use goodput and 99th-percentile FCT for mice flows ($\leq 100KB$) as the main performance metric. 

\subsection{Single-pod evaluation}
\label{sec-intra-pod-evaluation}

\begin{figure}[t]
    \centering
  
    \subfigure[Mice flow FCT with various $k$ for top-$k$.]{
      \includegraphics[width=0.18\textwidth]{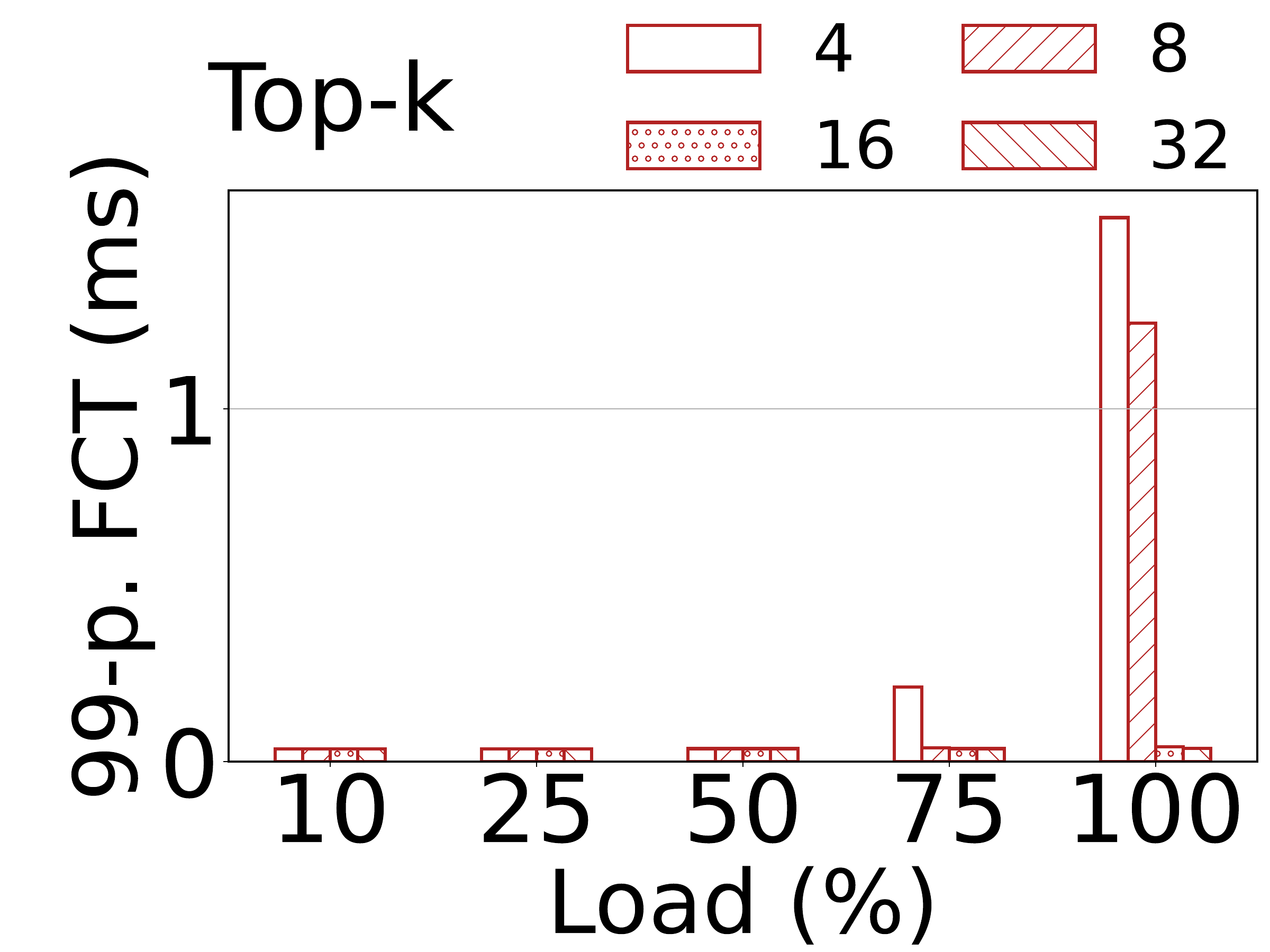}
      \label{topk_goodput_fct_singlepod}
 }
    \subfigure[CDF of mice flow FCT with optimistic sending on/off.]{
      \includegraphics[width=0.16\textwidth]{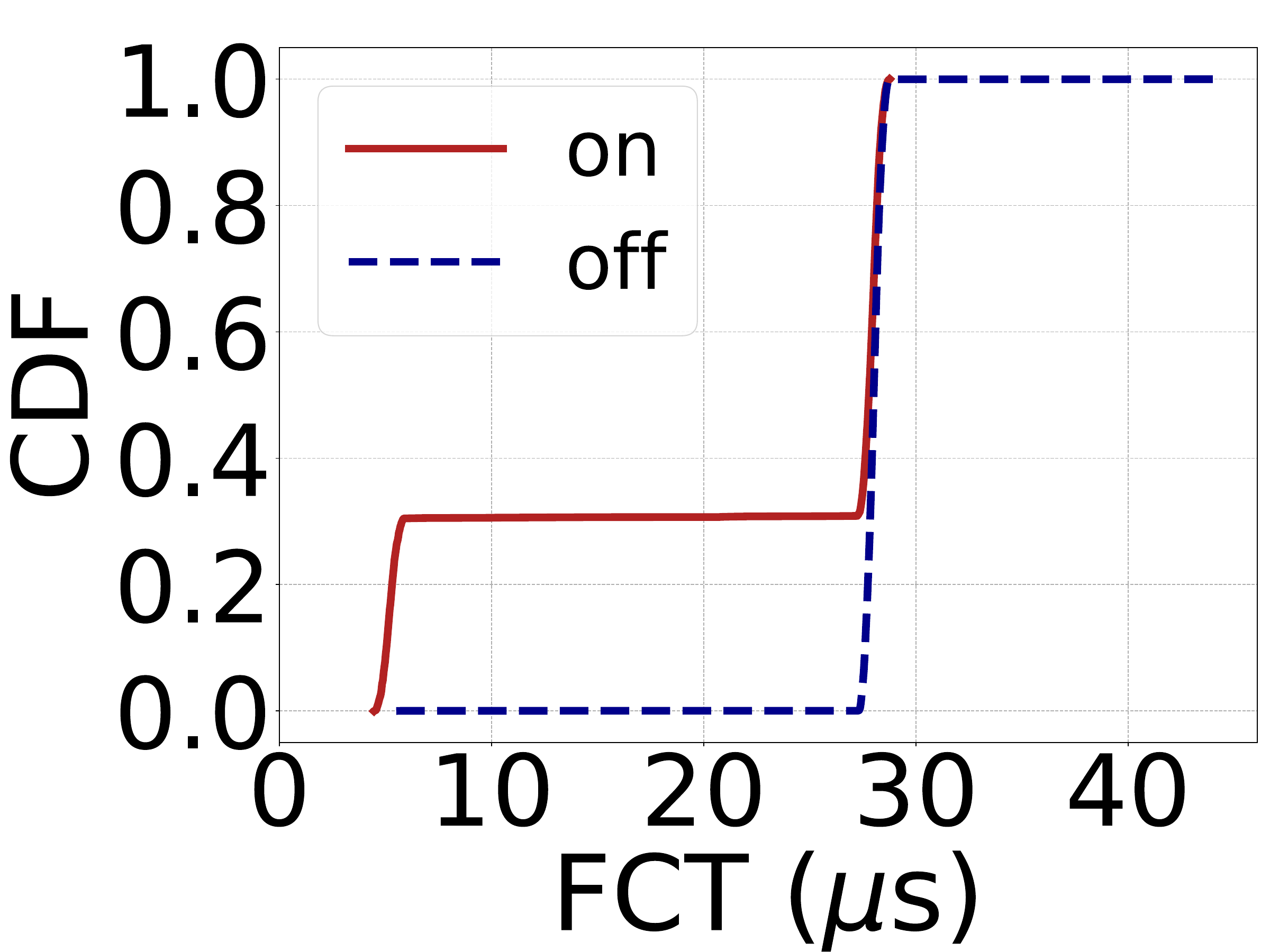}
      \label{cdf_optimistic_sending}
 }
  
    \vspace{-1pt}
  
    \subfigure[\hspace{-2pt}Incast finish time.]{
      \includegraphics[width=0.20\textwidth]{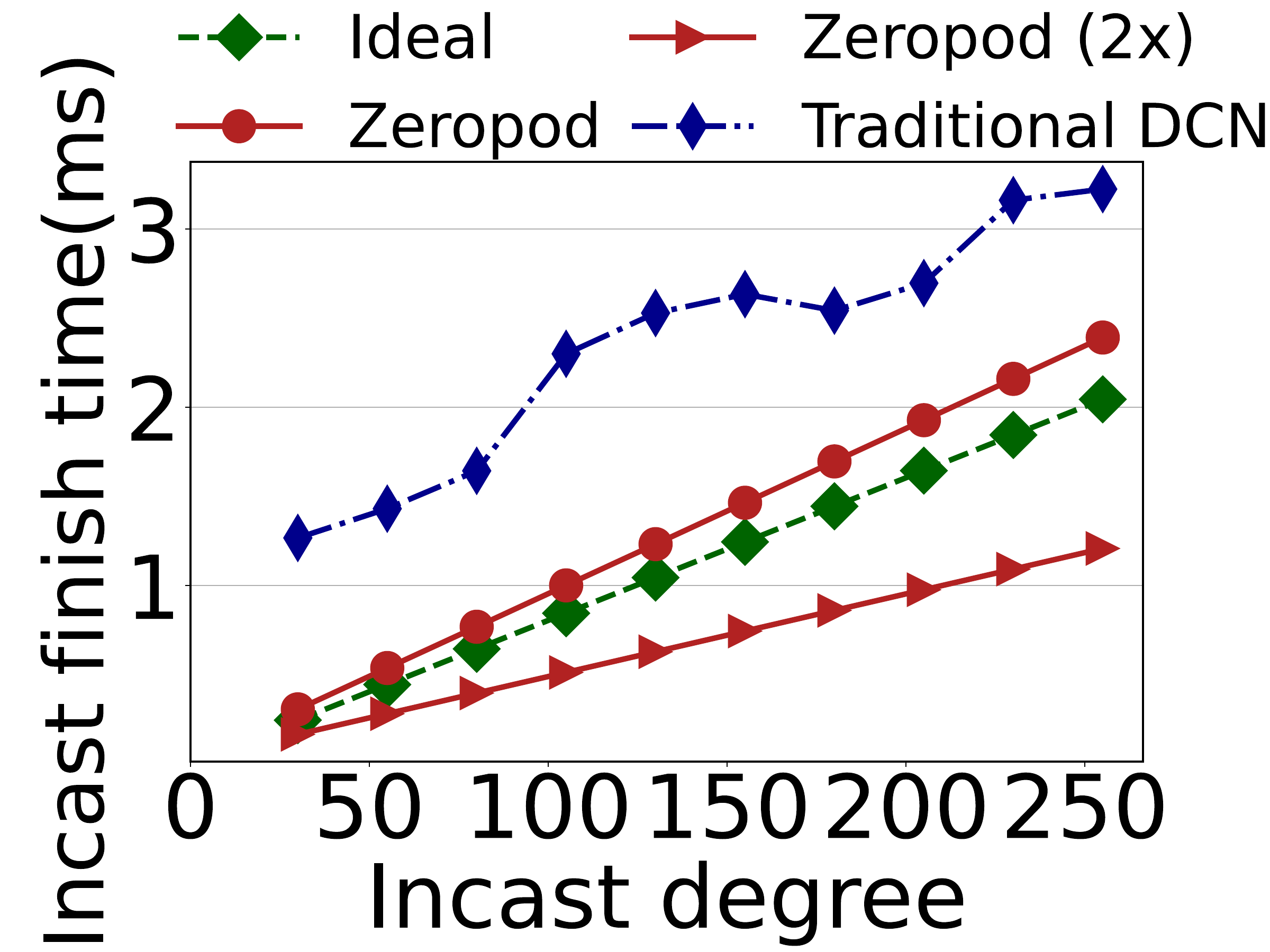}
      \label{incast_singlepod}
 }
    \subfigure[\hspace{-2pt}Single-pod collective communication traffic finish time (ms).]{
      \includegraphics[width=0.20\textwidth]{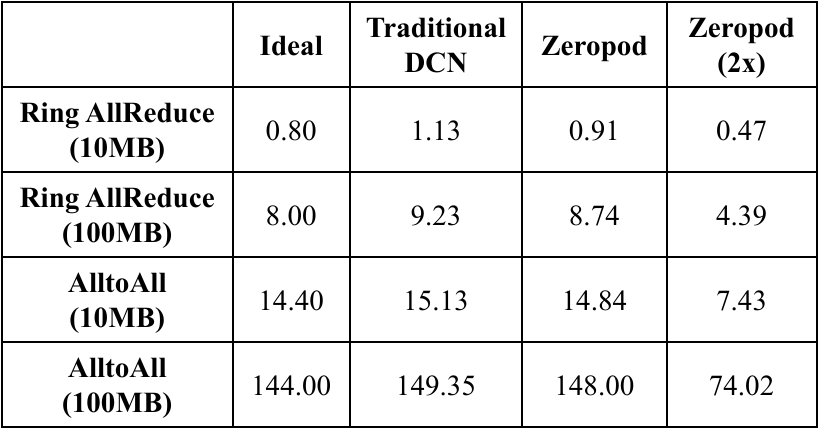}
      \label{dml_intra_pod}
 }
  
    \vspace{-6pt}
    \caption{Single-pod microbenchmarks.}
    \label{fig-intra-pod-microbenchmarks}
    \vspace{-6pt}
\end{figure}

\noindent\textbf{Top-$k$ mechanism for mice flow FCT reduction.} 
Here we investigate the optimal $k$ for mice flow prioritization that balances FCT and RTS information overhead.
We vary $k$ from 4 to 32. The goodput remains the same, so we omit the results. The FCT of mice flows decreases with a larger $k$, and remains stable when $k$ is larger than 16, as shown in Fig.~\ref{topk_goodput_fct_singlepod}. This indicates that a modest amount of flow information (i.e., 16) is sufficient for good performance. We use $k$=16 in later experiments.

\noindent \textbf{Optimistic sending mechanism for mice flow FCT reduction.} 
We verify the effectiveness of the optimistic sending mechanism under low loads. The network load is set to 10$\%$. 
Since we limit the optimistic sending mechanism to only taking effect on flows shorter than the threshold of 5 cells, we only report the FCT distribution of these mice flows ($<$ 3KB). Fig. \ref{cdf_optimistic_sending} shows that more than 30$\%$ of flows achieve much lower FCT with the mechanism on, and the tail latency of mice flows is significantly reduced, confirming its effectiveness.

\begin{figure}[t]
    \centering
  
    \begin{minipage}[h]{0.45\textwidth}
        \hspace{0.1in}\includegraphics[width=0.9\linewidth]{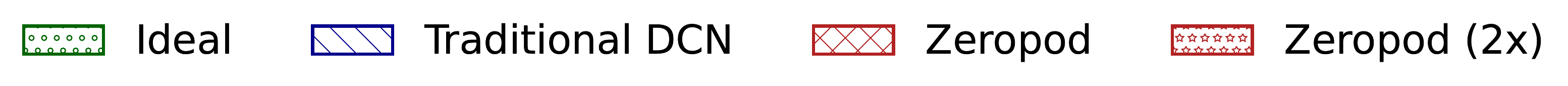}
        \vspace{-0.05in}
    \end{minipage}
  
    \subfigure[Goodput.]{
        \begin{minipage}[b]{0.18\textwidth}
          \centering
          \includegraphics[width=\linewidth]{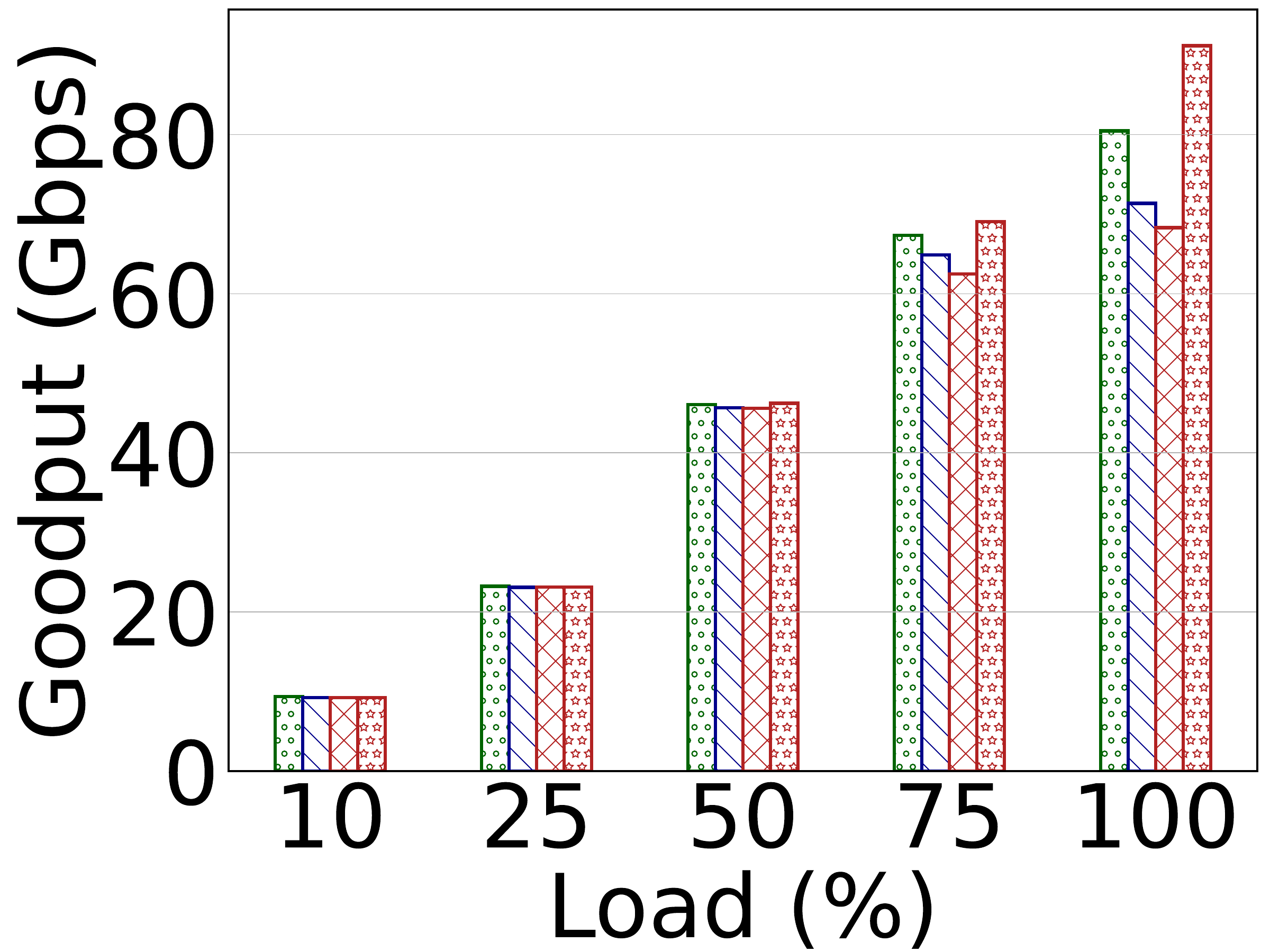}
        \label{Main_results_goodput_singlepod}
        \vspace{-0.1in}
        \end{minipage}
 }
    \subfigure[Mice flow FCT.]{
        \begin{minipage}[b]{0.18\textwidth}
          \centering
          \includegraphics[width=\linewidth]{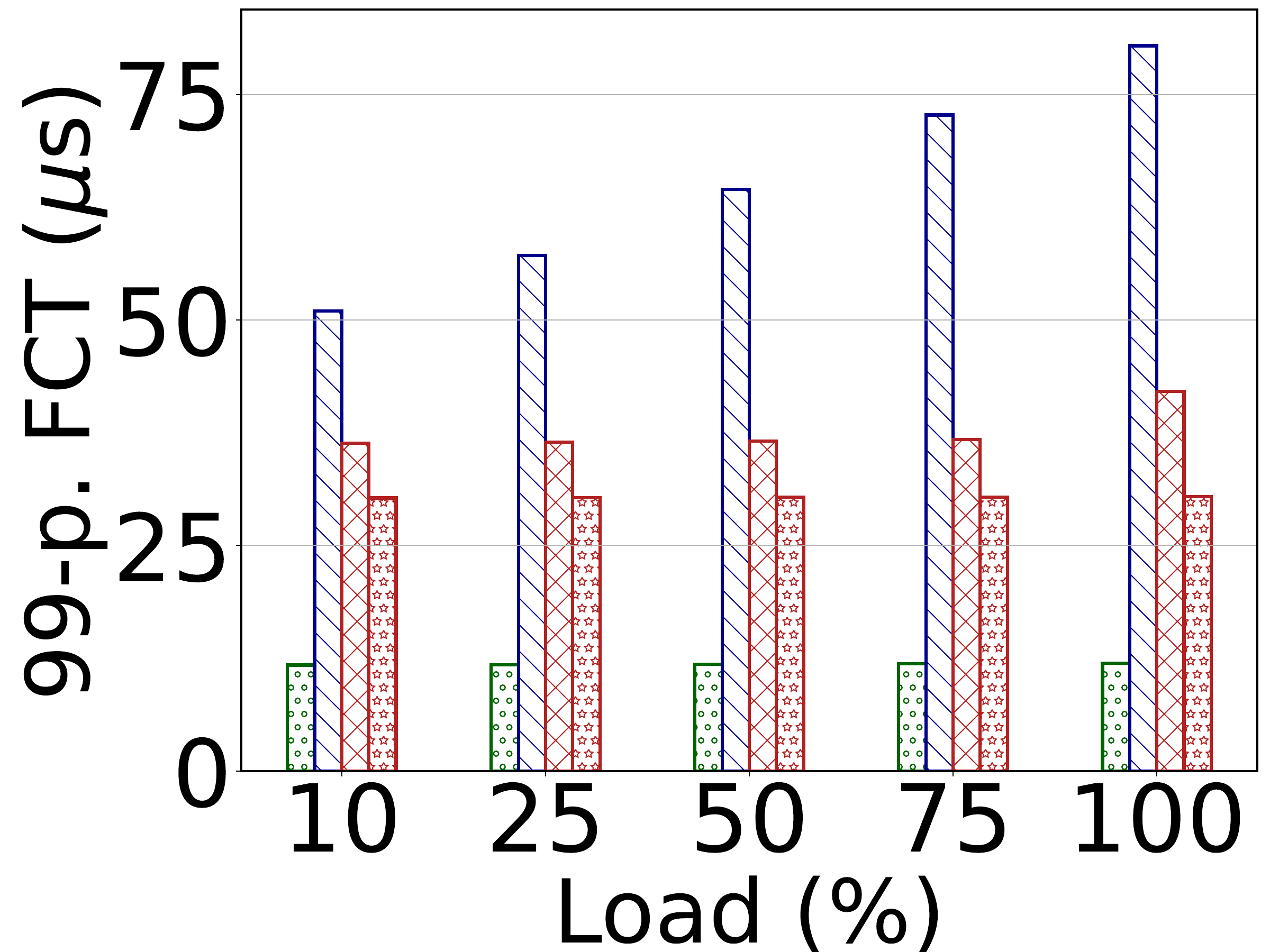}
        \label{Main_results_FCT99_singlepod}
        \vspace{-0.1in}
        \end{minipage}
 }
    \caption{Goodput and FCT at different single-pod loads.}
    \label{fig: overall performance in one pod}
    \vspace{-6pt}
\end{figure}

\noindent\textbf{Incast traffic and Collective Communication.} 
To test with diverse traffic patterns, we evaluate Zeropod under a pure incast workload, which is challenging for traditional DCNs, and also a typical collective communication workload in distributed machine learning (DML), and observe the finish time. 
We generate the incast by randomly selecting varying numbers of senders, each sending 100 KB to the same receiver simultaneously. As shown in Fig.~\ref{incast_singlepod}, by avoiding conflicts and buffer queueing or buffer overflow, the incast finish time of Zeropod is close to that of Ideal, and is much lower than the traditional DCN. 
Zeropod (2$\times$), with a larger bandwidth, can perform even better.

For DML, following its traffic characteristics \cite{10.1145/3663408.3663409}, we generate a ring AllReduce workload that involves all 256 endhosts in one pod forming a permutation, and an AlltoAll workload that involves 16 endhosts (1 from each rack) sending to each other in one pod, considering simulation speed limitation. Fig. \ref{dml_intra_pod} (the size is per-flow size) shows that Zeropod performs equally or better than the traditional DCN, and Zeropod (2$\times$) performs significantly better.

\noindent{\textbf{Overall performance.}} 
As shown in Fig. \ref{Main_results_goodput_singlepod}, Zeropod achieves a similar goodput with other schemes at load $\leq 75\%$, and a slightly lower goodput at 100$\%$ load. 
The lower goodput originates from the tradeoff we made, limiting the number of rounds for iterative scheduling to 6 in the evaluation setup to balance latency and goodput. However, when taking the increased switching capacity into account, Zeropod (2$\times$) can achieve a much higher goodput than the conventional DCN and approaches the ideal upper bound considering the header overhead, even at 100$\%$ load.

Fig. \ref{Main_results_FCT99_singlepod} shows that Zeropod delivers consistently lower tail FCT than the conventional DCN for mice flows, where FCT is reduced approximately by half at high loads. 
Zeropod (2$\times$) can further significantly reduce the FCT.

\subsection{Multi-pod evaluation}
\label{sec-inter-pod-evaluation}

\begin{figure}[t]
    \centering
  
    \begin{minipage}[h]{0.45\textwidth}
        \hspace{0.15in}\includegraphics[width=0.85\linewidth]{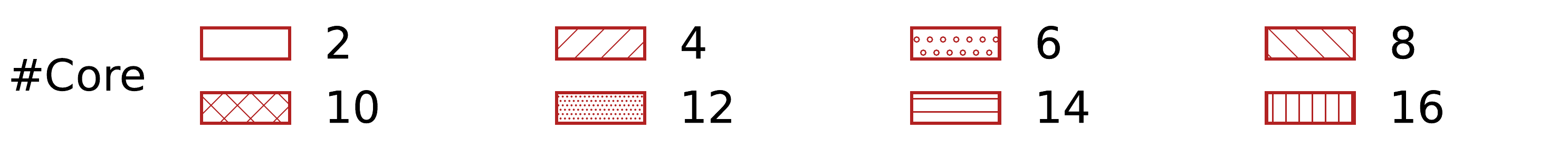}
        \vspace{-0.1in}
    \end{minipage}
  
    \subfigure[Goodput]{
        \begin{minipage}[b]{0.18\textwidth}
          \centering
          \includegraphics[width=\linewidth]{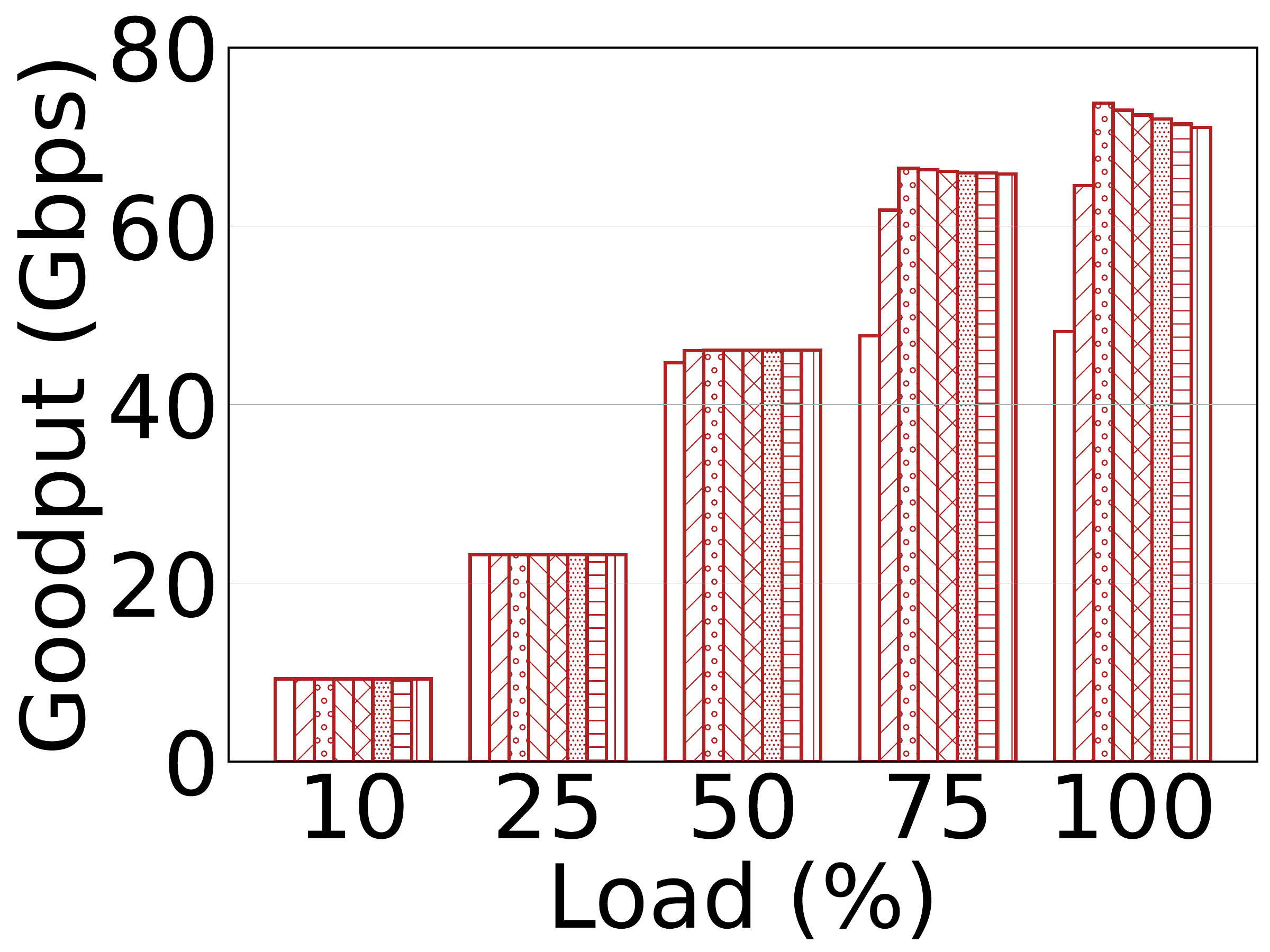}
        \label{inter_pod_core_switch_num_goodput}
        \vspace{-0.1in}
        \end{minipage}
 }
    \subfigure[Mice flow FCT]{
        \begin{minipage}[b]{0.18\textwidth}
          \centering
          \includegraphics[width=\linewidth]{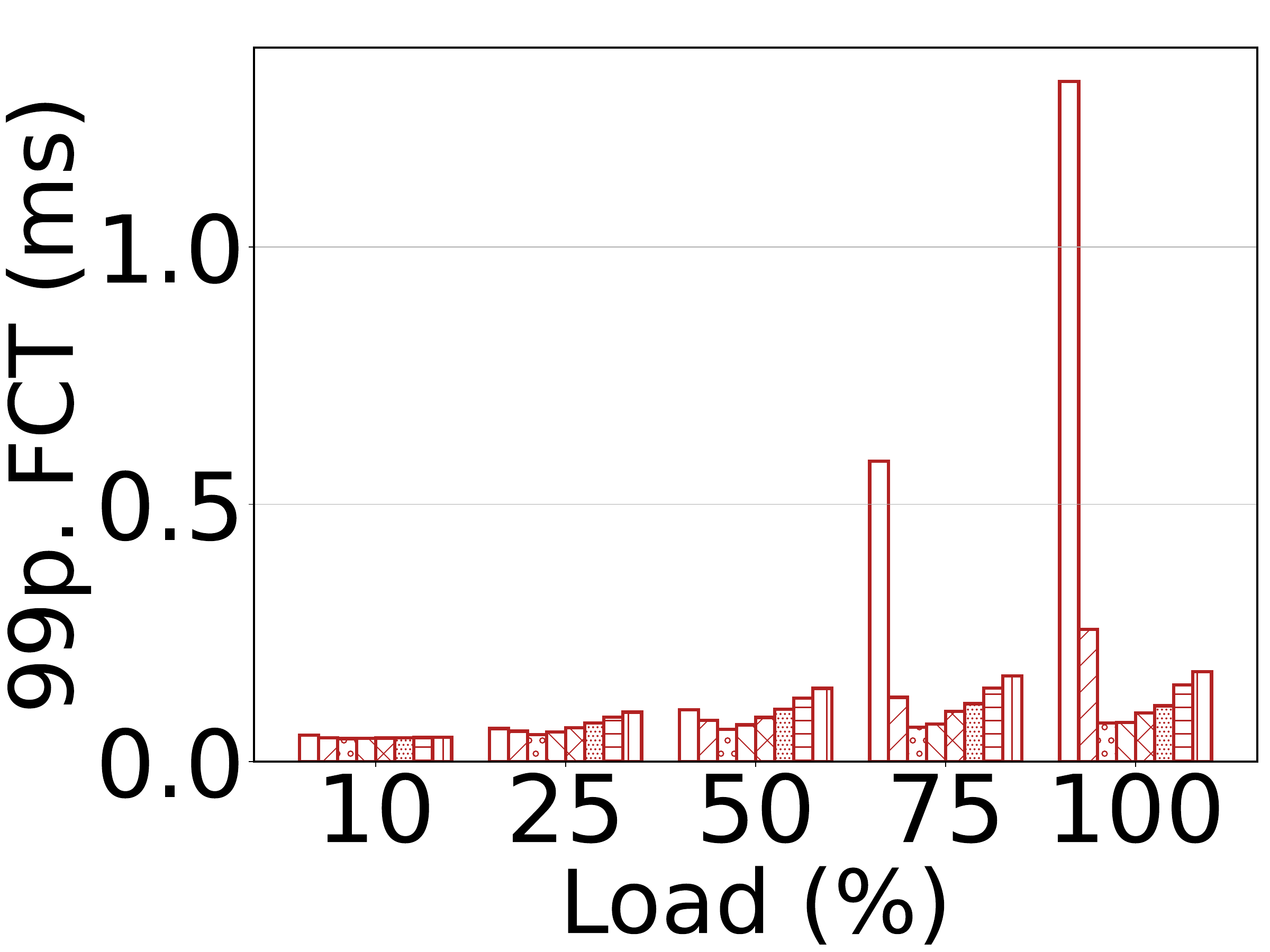}
        \label{inter_pod_core_switch_num_FCT}
        \vspace{-0.1in}
        \end{minipage}
 }
    \caption{Goodput and FCT with various numbers of Core switches.}
    \label{fig: overall performance under various number of Core switches}
    \vspace{-0.1in}
\end{figure}

\begin{figure}[t]
    \centering
  
    \begin{minipage}[b]{0.20\textwidth}
      \includegraphics[width=\linewidth]{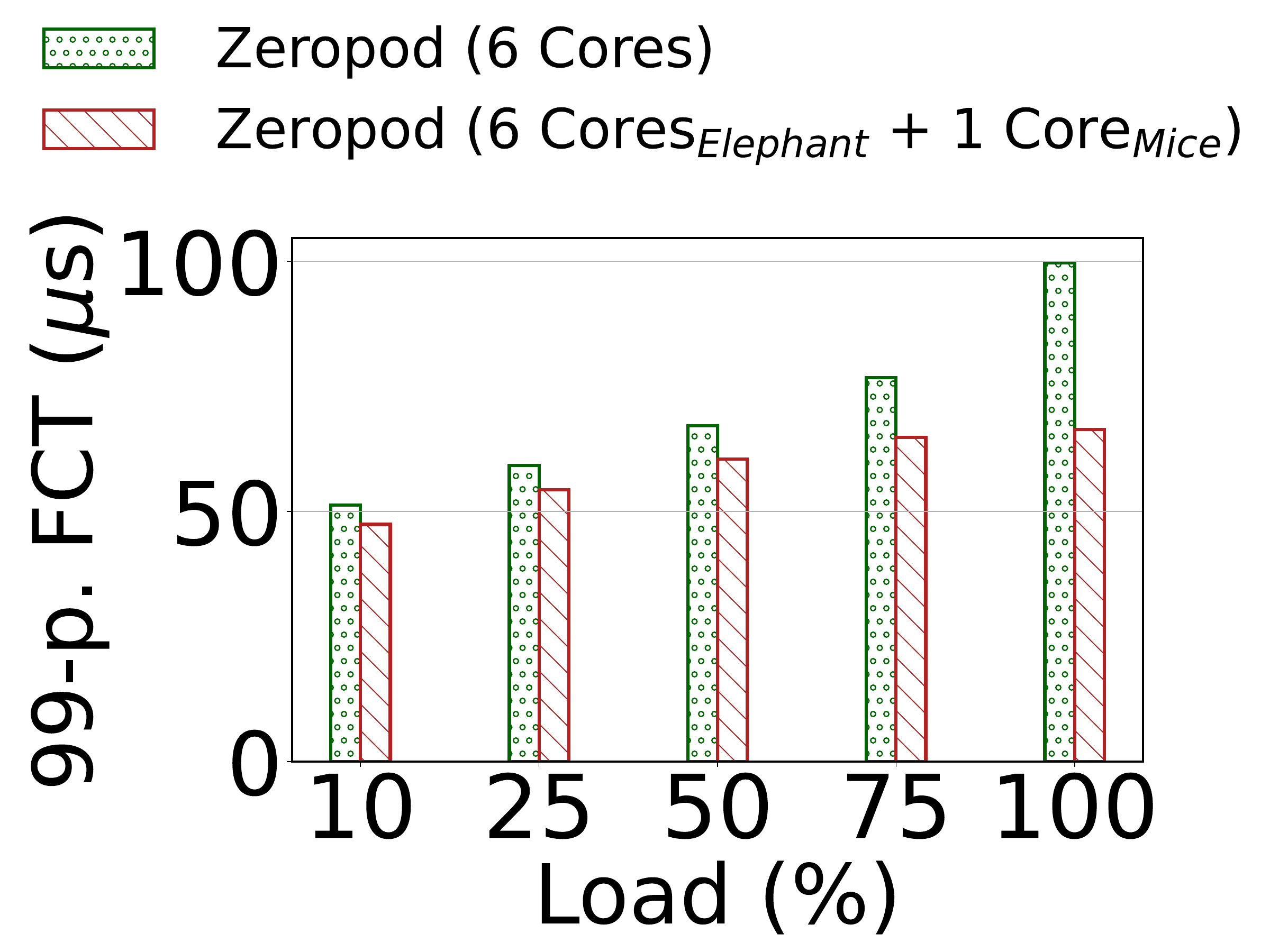}
    \vspace{-0.2in}
    \caption{Mice flow FCT with one extra Core for mice flow.}
    \label{inter_pod_one_more_core_FCT}
    \end{minipage}
    \begin{minipage}[b]{0.20\textwidth}
      \includegraphics[width=\linewidth]{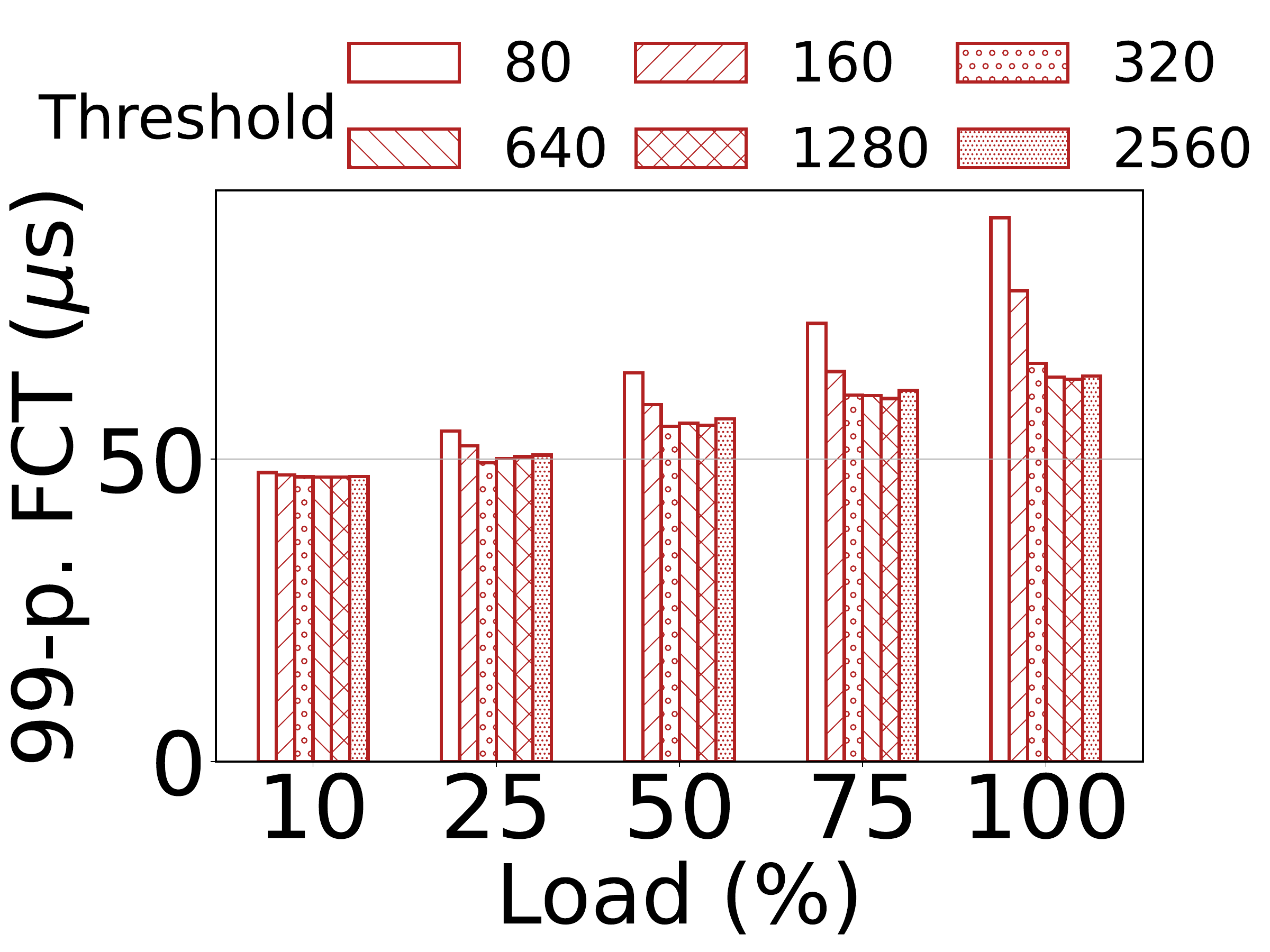}
    \vspace{-0.2in}
    \caption{Mice flow FCT with various backpressure thresholds (in \#data cells).}
    \label{inter_pod_backpressure_threshold_FCT}
    \end{minipage}
    \vspace{-0.2in}
\end{figure}

\begin{figure}[t]
    \centering
  
    \subfigure[Buffer occupancy of Core switches with no extra Core.]{
        \begin{minipage}[b]{0.19\textwidth}
          \centering
          \includegraphics[width=\linewidth]{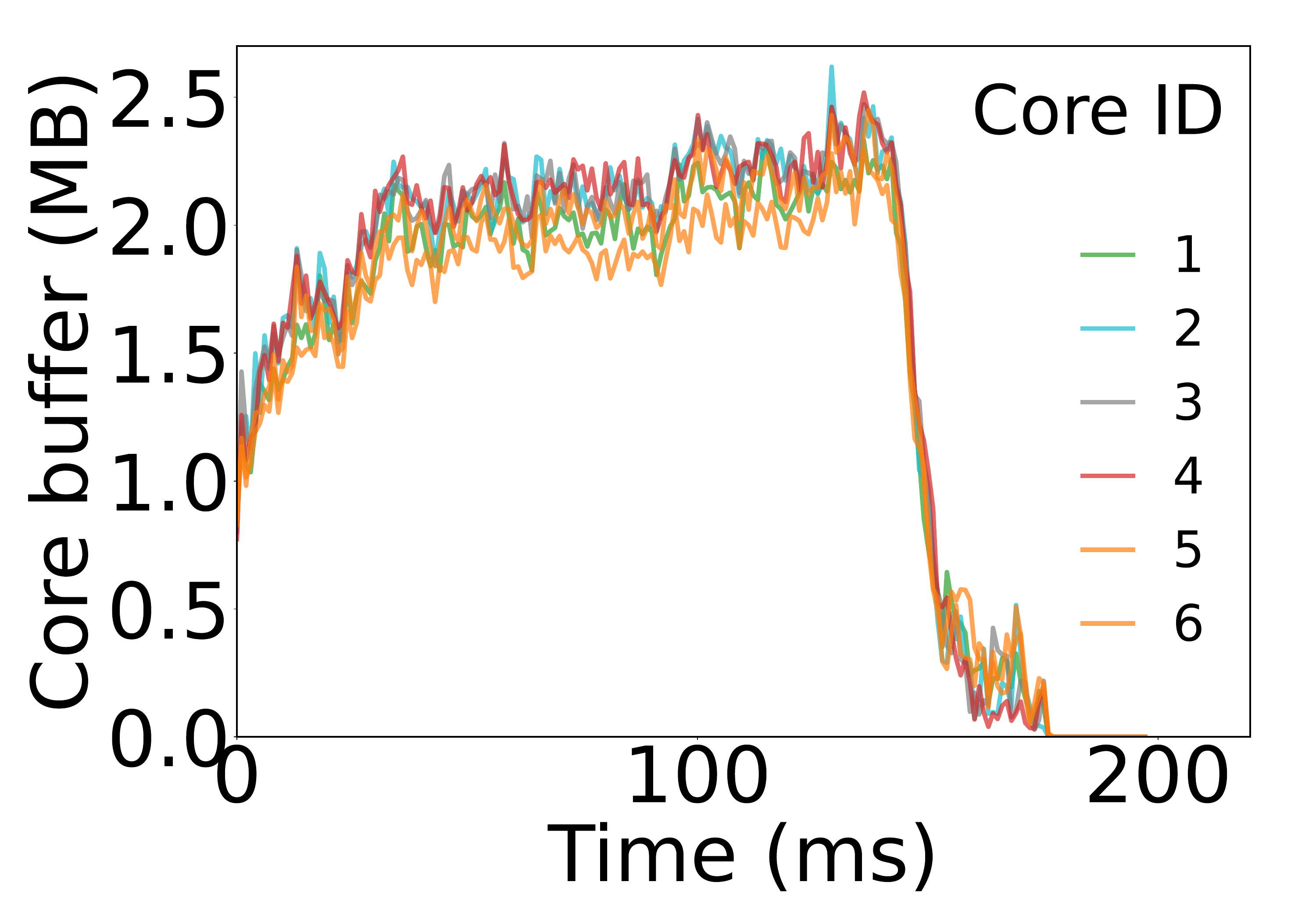}
        \label{inter_pod_core_buffer_occupancy_no_extra_core}
        \vspace{-0.1in}
        \end{minipage}
 }
    \subfigure[Buffer occupancy of Cores with one extra Core \#0.]{
        \begin{minipage}[b]{0.19\textwidth}
          \centering
          \includegraphics[width=\linewidth]{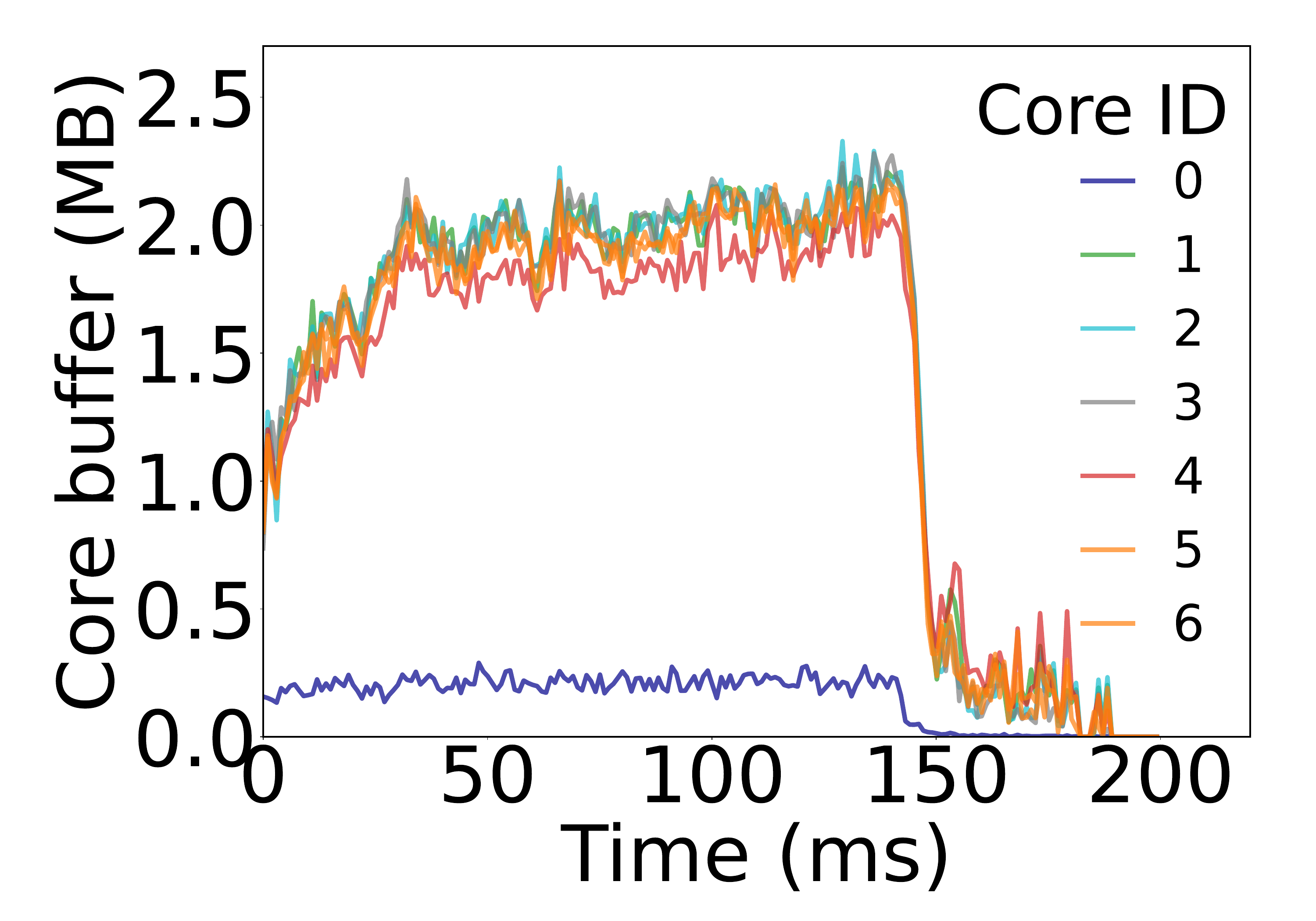}
        \label{inter_pod_core_buffer_occupancy_extra_core}
        \vspace{-0.1in}
        \end{minipage}
 }
    \vspace{-0.05in}
    \caption{Buffer occupancy of Core switches without/with one extra Core.}
    \vspace{-0.05in}
    \label{fig: overall performance under various number of Core switches buffer occupancy}
\end{figure}

\begin{figure}[t]
    \centering
  
    \begin{minipage}[b]{0.20\textwidth}
      \includegraphics[width=\linewidth]{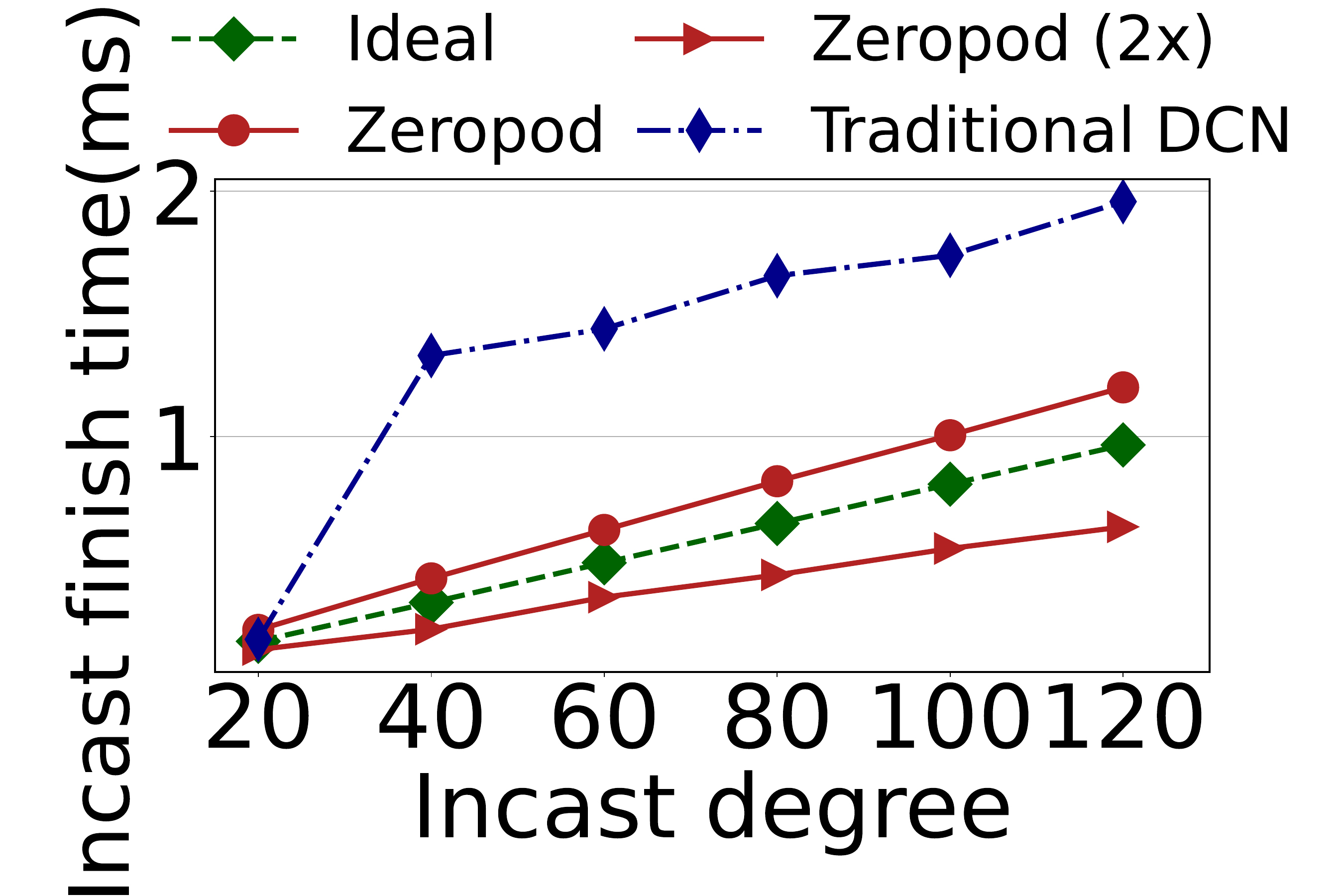}
    \vspace{-0.3in}
    \caption{Multi-pod incast finish time.}
    \label{inter_pod_incast}
    \end{minipage}
    \quad
    \begin{minipage}[t]{0.22\textwidth}
    \vspace{-0.9in}
      \includegraphics[width=\linewidth]{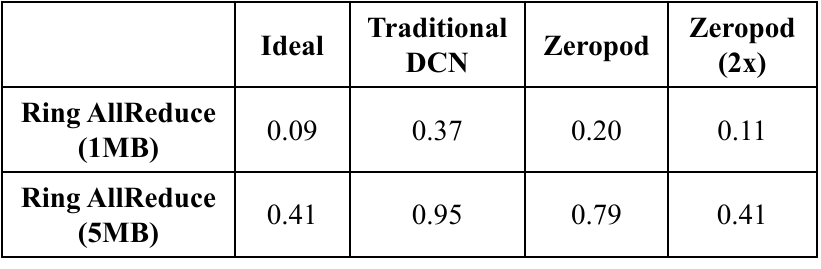}
    \vspace{-0.05in}
    \caption{Multi-pod collective communication traffic finish time (ms).}
    \label{dml_inter_pod}
    
    \end{minipage}
    \label{fig-micro-inter}
\end{figure}

\noindent \textbf{Number of Core switches.} 
Like traditional DCNs, Zeropod introduces oversubscription between pods due to traffic locality.
We investigate what the minimal number of Core switches is to achieve good performance and save cost. As shown in Fig.~\ref{inter_pod_core_switch_num_goodput} and Fig.~\ref{inter_pod_core_switch_num_FCT}, extremely low numbers of Core switches will lead to a significant performance drop. When the number increases to 6, both goodput and FCT reach a good and stable state. Following our design in \S\ref{sec-design-0-1}, we further add one extra Core for mice flows and show the result in Fig.~\ref{inter_pod_one_more_core_FCT}.
With just one mice flow exclusive Core, the FCT of mice flows is reduced by more than 30\% at 100\% load. We thus set the number of Cores to 6 and add one extra Core for mice flows.

\noindent \textbf{Backpressure threshold of Core switches, and their buffer occupancy.} 
Backpressure threshold controls the queue length in Core switches. 
A high threshold can result in long queues in Core switches and incur long queuing delay or even buffer overflow. A low threshold will bring down bandwidth utilization and, in turn, damage FCT because the paused Core port will drain its buffer when waiting for the enabling signal to take effect. Fig.~\ref{inter_pod_backpressure_threshold_FCT} shows that 320 data cells can achieve good FCT in our setup.

With this setup, we also monitor the buffer occupancy of Core switches with and without the extra Core in Fig.~\ref{inter_pod_core_buffer_occupancy_no_extra_core} and Fig.~\ref{inter_pod_core_buffer_occupancy_extra_core}. Thanks to the backpressure mechanism, Core's buffer requirement is low (several MBs) compared with commodity switches \cite{wheeler2019tomahawk} (tens of MBs), improving its practicality.

\noindent\textbf{Incast traffic and Collective Communication.} Similar to single-pod, we test Zeropod under multi-pod incast and DML collective communication workloads. Note that we do not test the AlltoAll workload here because it is usually limited to one pod scale \cite{deepseekai2025deepseekv3technicalreport}. Fig.~\ref{inter_pod_incast} and Fig.~\ref{dml_inter_pod} show similar trends to single-pod, confirming the effectiveness of Zeropod.

\begin{figure}[t]
    \centering
    \begin{minipage}[h]{0.40\textwidth}
    \includegraphics[width=\linewidth]{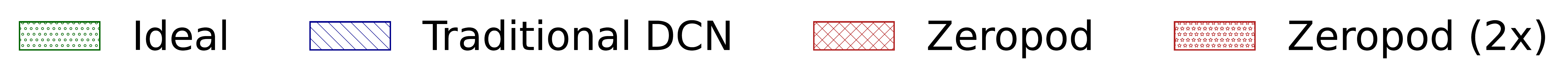}
        \vspace{-0.3in}
    \end{minipage}
  
    \subfigure[Goodput]{
        \begin{minipage}[b]{0.18\textwidth}
          \centering
          \includegraphics[width=\linewidth]{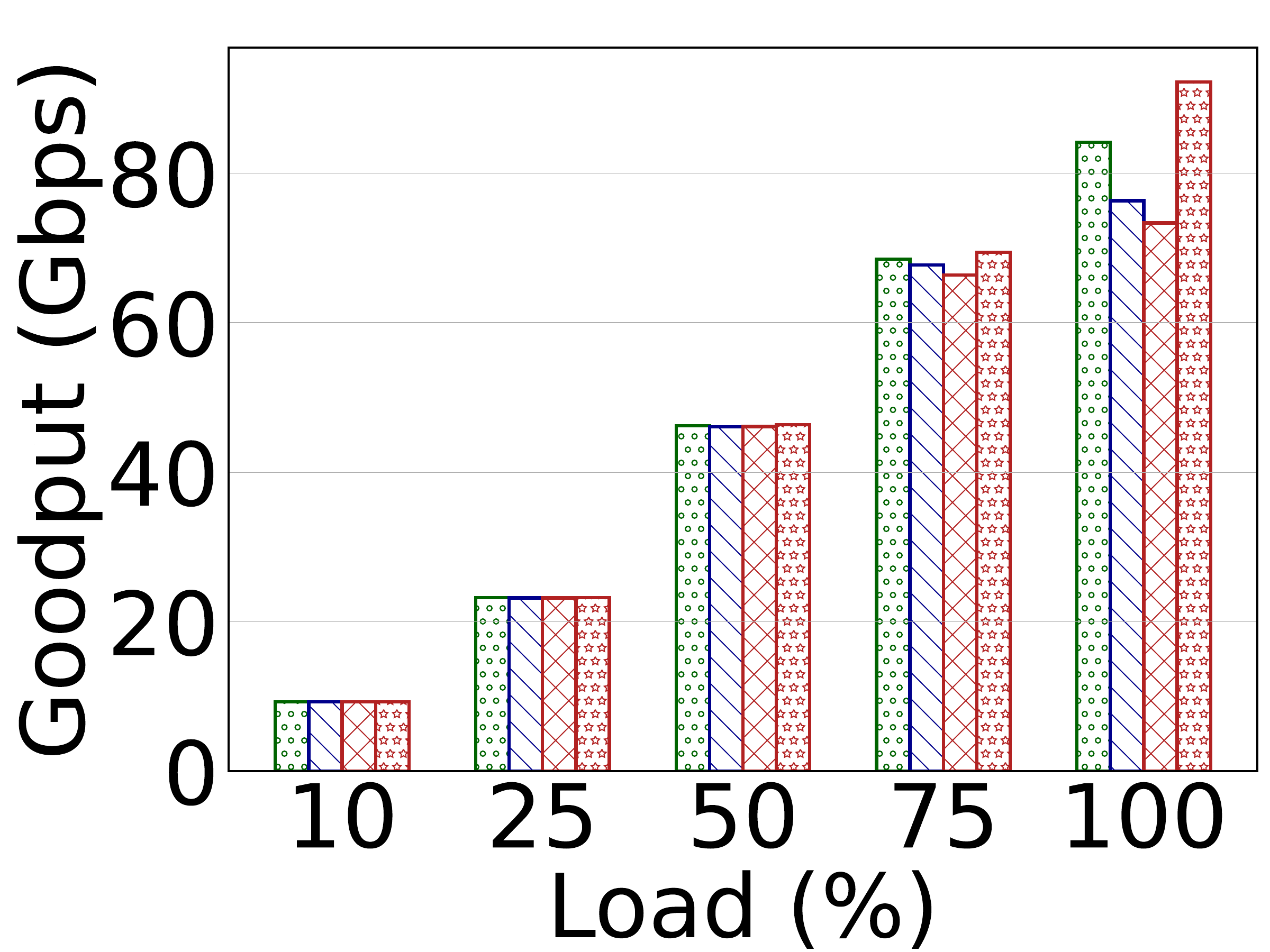}
        \label{Main_results_goodput}
        \vspace{-0.1in}
        \end{minipage}
 }
    \subfigure[Mice flow FCT]{
        \begin{minipage}[b]{0.18\textwidth}
          \centering
          \includegraphics[width=\linewidth]{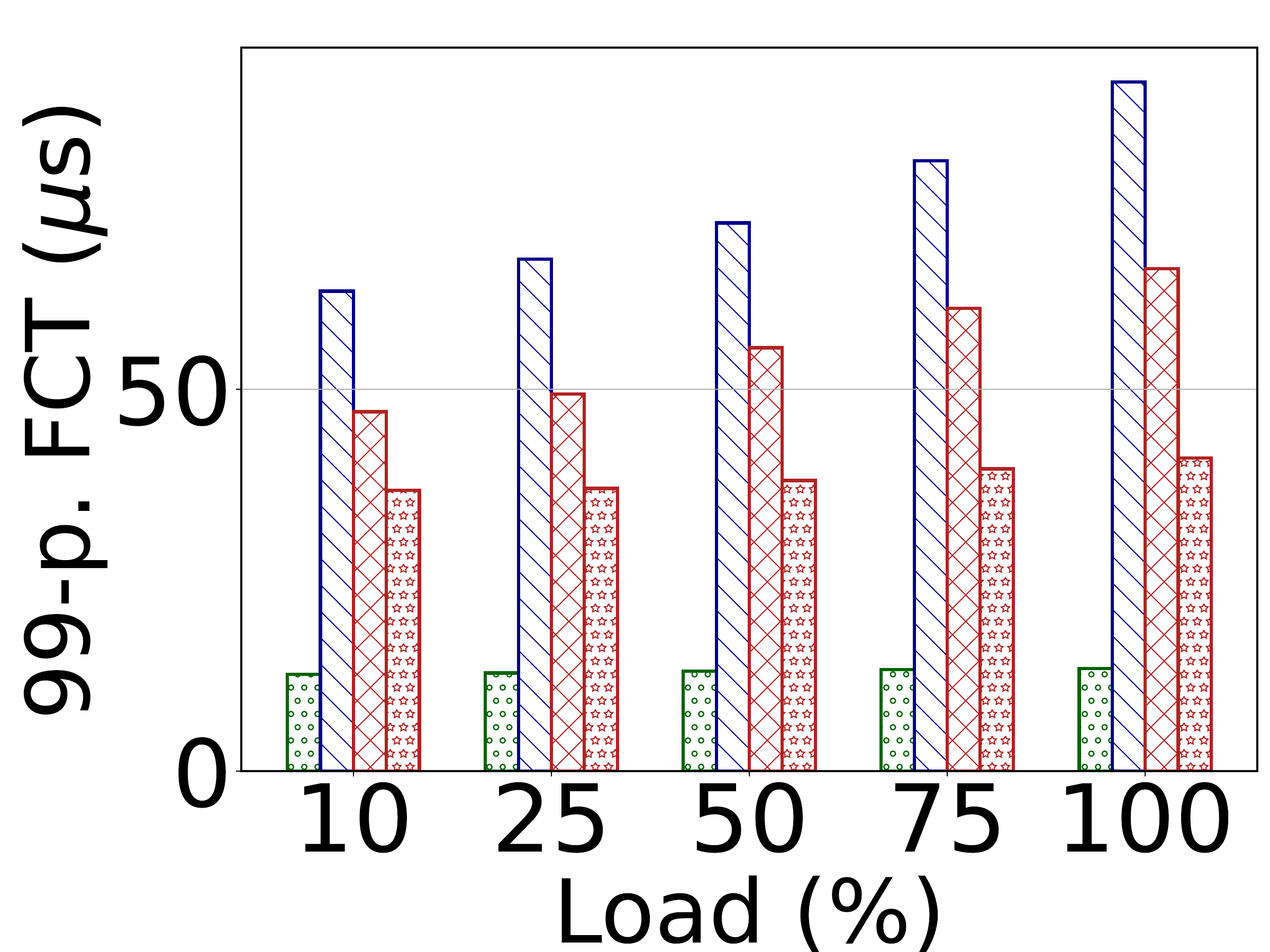}
        \label{Main_results_FCT99}
        \vspace{-0.1in}
        \end{minipage}
 }
    \vspace{-6pt}
    \caption{Goodput and FCT at different multi-pod loads.}
    \label{fig: overall performance multi pods}
    \vspace{-6pt}
\end{figure}

\noindent{\textbf{Overall performance.}} 
Fig. \ref{fig: overall performance multi pods} shows that the multi-pod performance of Zeropod exhibits similar trends to that of single-pod. Notably, even though inter-pod traffic has to experience two stages of transmission, with optimizations like the exclusive Core switch for mice flows, Zeropod still achieves significantly lower tail FCT than the traditional DCN, showing its effectiveness in multi-pod scenarios.

\section{Related Work}

\noindent\textbf{Enhancing traffic control in existing DCNs.}
In the post Moore's Law era, the switching capacity and buffer size of packet switches are becoming insufficient to fulfill the FCT and goodput needs for emerging traffic \cite{ballani2020sirius, 295507, bfc2022nsdi}. One line of work enhances the traffic control in existing DCN architectures \cite{pyrrha, pred, li2019hpcc, bfc2022nsdi,bolt2023nsdi, 295507}. Among them, HPCC \cite{li2019hpcc} uses network telemetry information provided by programmable switches to enable precise congestion control, and Bolt \cite{bolt2023nsdi} uses programmable switches to control traffic at a per-hop granularity. 
Although they achieve more efficient usage of the small switch buffers, they do not ease the burden of switching chips. They may even make the chips more complex with the advanced control logic, further hindering the sustainable development of DCNs. In contrast, Zeropod rethinks DCN architecture and proposes zero-buffer packet switching, which simplifies the chip while preserving the flexibility of packet switching, enabling higher switching capacity under the same-generation chip manufacturing technology.

\noindent\textbf{Simplified DCN architecture.}
Another line of work \cite{zilberman2019stardust, jin2016your, firestone2018azure, li2018dumbnet, 10.1145/2486001.2491722, 10.1145/285237.285273} simplifies traditional networks by pushing the functionalities to the edges. However, they are still based on the traditional buffered packet switching. Zeropod follows this trend, presenting a zero-buffer packet-switching approach with centralized control, pushing the simplified packet-switched network to an end. 
Compared with recent circuit switching~\cite{shrivastav2019shoal,mellette2020expanding,ballani2020sirius,10.1145/3651890.3672273} works, which also have no in-network buffer, Zeropod eliminates the need for path-level resource reservation, enabling flexible link-level hop-by-hop scheduling and thus potentially higher goodput.

There are also DCNs using centralized schedulers, like Fastpass \cite{perry2014fastpass}. However, Fastpass suffers from limited scalability because one scheduler needs to control the whole DCN-scale network with thousands of endhosts \cite{roy2015inside}. Meanwhile, Fastpass only achieves zero queue instead of zero buffer since in-band control packets still need to be buffered, limiting the simplicity of its switching chips. In contrast, Zeropod highlights its scalability. It limits the centralized-controlled domain to each pod, where multiple schedulers are used, and each scheduler's work is effectively reduced. Multiple zero-buffer pods are distributedly coordinated with buffered Core switches, expanding the network to the DCN scale.

\section{Conclusion}
In this paper, we present Zeropod, a new DCN architecture with zero-buffer packet switching. 
With a centralized scheduler, buffers are removed inside pods to extremely simplify the data plane.
Multiple pods are connected by buffered Core switches to ease the complexity of scheduling, thus expanding the design of Zeropod to a large-scale DCN.
Zeropod explores an extreme end of the design spectrum, and we hope it can encourage further exploration in the DCN community.

\bibliographystyle{IEEEtran}
\bibliography{IEEEabrv, reference}

\end{document}